\documentclass[final,5p,times,twocolumn,authoryear]{elsarticle}

\usepackage{amssymb}
\usepackage[T1]{fontenc}
\usepackage[utf8]{inputenc}
\usepackage{xcolor}
\usepackage{amsmath}
\usepackage{url}
\usepackage{hyperref} 

\usepackage{placeins}

\graphicspath{{Figs/}{SI_Figs/}}

\begin{document}

\begin{frontmatter}



\title{Surface-Dependent Phonon Dynamics in 9-Armchair Graphene Nanoribbon Arrays}


\author[empaTNI,baselPhys,baselSNI]{Ángel Labordet Álvarez\corref{cor1}}

\author[empaNano]{Gabriela Borin Barin}

\author[empaTNI,baselPhys,baselSNI]{Michel Calame\corref{cor1}}

\author[empaTNI,UZH]{Mirjana Dimitrievska\corref{cor1}}

\cortext[cor1]{%
Corresponding authors:
\href{mailto:angel.labordet@empa.ch}{angel.labordet@empa.ch} (Ángel Labordet),
\href{mailto:michel.calame@empa.ch}{michel.calame@empa.ch} (Michel Calame),
\href{mailto:mirjana.dimitrievska@empa.ch}{mirjana.dimitrievska@empa.ch} (Mirjana Dimitrievska).
}

\affiliation[empaTNI]{%
  organization={Transport at Nanoscale Interfaces,
                Swiss Federal Laboratories for Materials Science and Technology (Empa)},
  city={Dübendorf},
  postcode={8600},
  country={Switzerland}
}

\affiliation[baselPhys]{%
  organization={Department of Physics, University of Basel},
  city={Basel},
  postcode={4056},
  country={Switzerland}
}

\affiliation[baselSNI]{%
  organization={Swiss Nanoscience Institute, University of Basel},
  city={Basel},
  postcode={4056},
  country={Switzerland}
}

\affiliation[empaNano]{%
  organization={nanotech@surfaces Laboratory,
                Swiss Federal Laboratories for Materials Science and Technology (Empa)},
  city={Dübendorf},
  postcode={8600},
  country={Switzerland}
}

\affiliation[UZH]{%
  organization={Department of Chemistry, University of Z\"urich},
  city={Z\"urich},
  postcode={8057},
  country={Switzerland}
}

\begin{abstract}
Atomically precise graphene nanoribbons (GNRs) have electronic properties tunable through their width and edge structure, making them promising building blocks for nanoscale electronics and optoelectronics. Their integration into devices also requires understanding how the supporting surface and neighbouring ribbons affect their vibrations. Here, we use temperature-dependent Raman spectroscopy to compare five configurations of 9-armchair GNR arrays differing in substrate, alignment and coverage. Transferring the same unaligned film from Au to a Raman-optimised Al$_2$O$_3$-coated support reduces the measured $D$- and $G$-mode redshift rates by about 80\%. This reduction is consistent with a substantial thermoelastic contribution. The fitted strain and anharmonic contributions depend on the assumed thermal expansion and strain transfer. Compared with the low-coverage array, the dense aligned Au array has higher extrapolated reference frequencies ($\omega_0$). On first heating, its $G$-mode linewidth decreases and then increases, forming an unusual intermediate-temperature minimum. The $D$-mode linewidth shows a similar, less pronounced variation. These findings establish substrate choice and array arrangement as routes to engineer nanoribbon vibrations, extending control beyond the atomic structure defined during synthesis.

\end{abstract}




\end{frontmatter}




\section{Introduction}
\label{introduction}

Atomically precise armchair graphene nanoribbons (AGNRs)
are one-dimensional semiconductors whose width and edge structure determine
their electronic and vibrational properties. Bottom-up synthesis on Au
produces chemically defined ribbons \cite{Cai2010BottomUpGNR,DiGiovannantonio2018GrowthDynamics},
and Raman spectroscopy provides non-destructive access to the $G$ band,
the confinement-activated $D$ band, and the radial-breathing-like mode
(RBLM) \cite{Saito2010RamanRibbons,Gillen2009VibrationalNanoribbons,
Liu2020InPlaneBreathing,Sheremetyeva2024ResonantRaman}. Confinement and armchair-edge symmetry activate the $D$ feature through zone folding, whereas the RBLM is a collective transverse oscillation whose frequency depends on ribbon width \cite{Nascimento2025PhononAssignment}. These signatures enable quality control before and after transfer to device-compatible substrates
\cite{BorinBarin2019SwitchingDevices,Overbeck2019OptimizedSubstrates,Zhang2026GNRNanoelectronics}. The electronic band gap of an AGNR is width dependent, which makes atomically precise ribbons a route from gapless graphene to semiconducting channels \cite{Geim2007RiseGraphene,Son2006EnergyGaps,Houtsma2021AtomicallyPrecise}.

Vicinal Au(788) consists of Au(111) terraces that are 16 atomic
rows wide, corresponding to $3.83$~nm in the ideal geometry, and
separated by steps one atomic layer high
\cite{Rousset2003VicinalSurfaces}. This periodic structure aligns
AGNRs along one in-plane direction
\cite{Linden2012AlignedGNRs}. Changing the precursor dose produces
either one ribbon row per terrace or a close-packed monolayer with
three rows per terrace
\cite{Darawish2025PrecursorCoverage}. Growth on Au(111), in contrast,
yields an unaligned array \cite{DiGiovannantonio2018GrowthDynamics}.
After synthesis, the aligned and unaligned films can be transferred by
different wet-transfer approaches
\cite{Zhang2026GNRNanoelectronics}
onto a Raman-optimised Al$_2$O$_3$/metal/SiO$_2$/Si substrate
\cite{Overbeck2019OptimizedSubstrates}. This sample matrix varies
substrate, alignment, and coverage.

Temperature-dependent Raman spectroscopy provides two distinct observables. The peak position gives the phonon frequency and shifts because of intrinsic phonon--phonon interactions and strain generated by thermal-expansion mismatch between the ribbon and substrate. The linewidth reflects phonon damping and dephasing, together with instrumental and inhomogeneous broadening. Uniform strain shifts the peak position without broadening it, whereas non-uniform strain can affect both the peak position and linewidth \cite{Yoon2011NegativeThermalExpansion,Bonini2007PhononAnharmonicities,Linas2015RamanThermalExpansion}. Carbon nanotubes provide a one-dimensional comparison for narrow AGNRs: their radial breathing and $G$ modes both evolve with temperature \cite{Raravikar2002RadialBreathing,Huang1998NanotubeRaman}. For atomically precise GNRs, Guo \emph{et al.} measured 7-AGNRs on Au(111) from 80 to 520~K, obtained $d\omega_G/dT=-0.026~\mathrm{cm}^{-1}\mathrm{K}^{-1}$, and associated the nonlinear RBLM response with a quartic channel \cite{Guo2022PhononAnharmonicities}. The present study compares temperature-dependent Raman responses of 9-AGNR arrays across configurations spanning different substrates, ribbon alignments, and packing densities.

Here we measure the RBLM, $D$, and $G$ modes of five 9-AGNR array configurations from 70 to 300~K. Peak positions and full widths at half maximum are obtained from full-window global Lorentzian fits. We model the temperature-dependent shifts of the RBLM, $D$, and $G$ Raman peak positions as the sum of a substrate--ribbon thermal-expansion-mismatch contribution and a lowest-order Klemens term describing three-phonon interactions, with all model assumptions stated explicitly. The analysis adds aligned and unaligned arrays, low- and high-coverage, Au and Raman-optimised substrates, thermal cycling, and mode-resolved linewidths to the one-dimensional GNR benchmark established by Guo \emph{et al.}

\begin{figure*}[!t] 
  \centering
  \includegraphics[width=\textwidth]{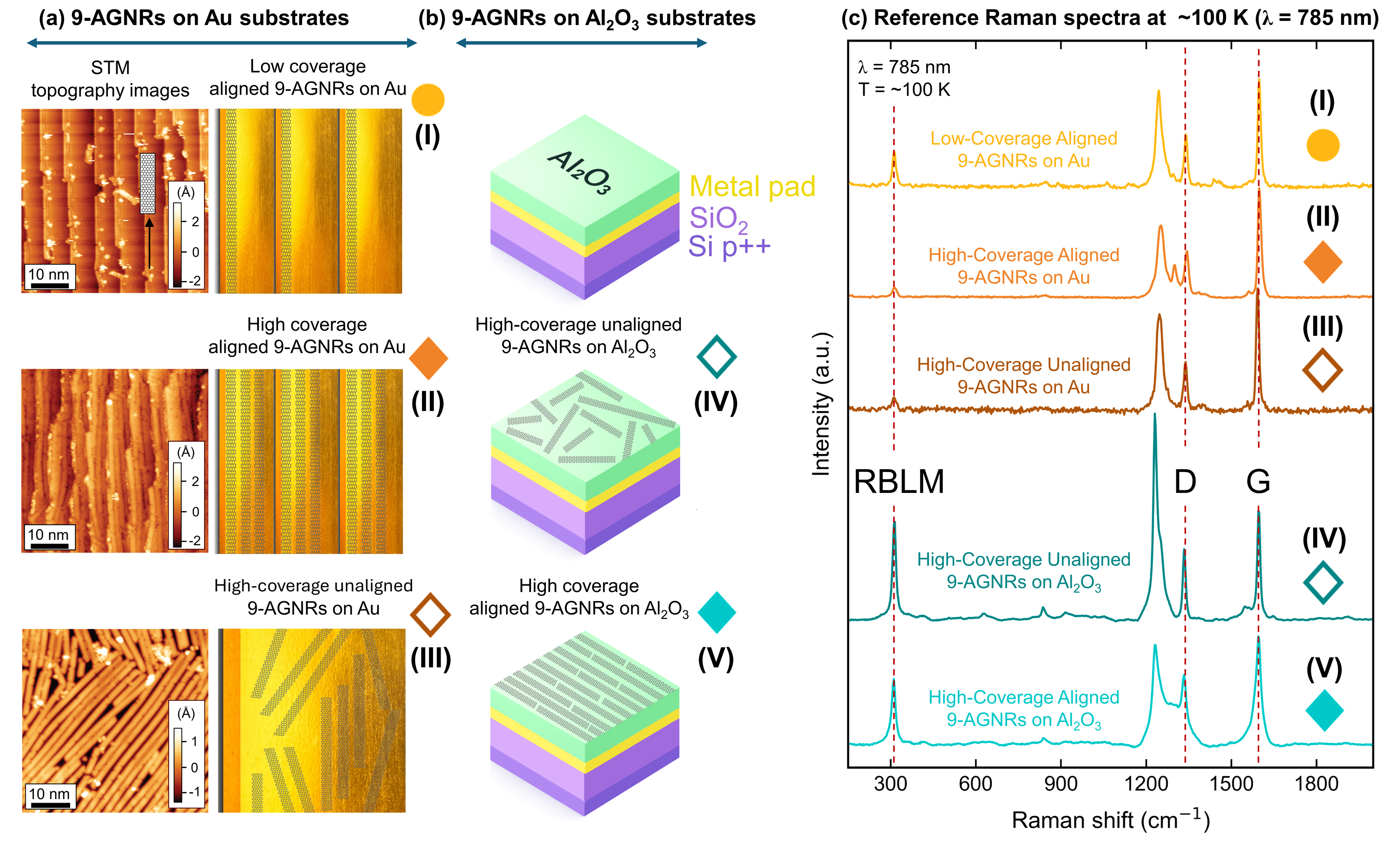}
  \caption{%
    Sample matrix of the five 9-AGNR array configurations.
    (a) Scanning tunnelling microscopy (STM) topography images (left) and schematic
    representations (right) of
    (I) \emph{Aligned Au, low-coverage}, (II) \emph{Aligned Au, high-coverage}, and
    (III) \emph{Unaligned Au, high-coverage}. (b) Cross-section of the Raman-optimised (RO)
    Al$_2$O$_3$/metal/SiO$_2$/Si stack and schematic representations of
    (IV) \emph{Unaligned RO, high-coverage}, and (V) \emph{Aligned RO, high-coverage}.
    (c) Raman spectra at 100~K for I, II, and V, 95~K for III, and 105~K for IV
    ($\lambda=785$~nm), normalised to the maximum in the $G$ window and
    highlighting the RBLM, $D$, and $G$ modes. Colours and symbols identify each configuration
    and are used throughout the manuscript.
  }
  \label{fig:Sample_Matrix}
\end{figure*}

\section{Methods}
\label{sec:methods}

\subsection{On-surface synthesis of 9-AGNR arrays}
\label{sec:synthesis}

9-AGNRs were synthesised by surface-assisted polymerisation followed by cyclodehydrogenation
using the automated GNR-reactor protocol established at Empa Swiss Federal Laboratories for Materials Science and Technology
\cite{BorinBarin2019SwitchingDevices}. Au substrates were cleaned by repeated
Ar$^+$ sputter/anneal cycles, the molecular precursor was deposited on room-temperature Au,
and sequential annealing at $200\,^{\circ}$C and $400\,^{\circ}$C produced
atomically precise 9-AGNRs. The reactor and complete recipe are reported in
Ref.~\cite{BorinBarin2019SwitchingDevices}.

Au(788) consists of Au(111) terraces that are 16 atomic rows wide, corresponding to $3.83$~nm in the ideal geometry. Adjacent terraces are separated by steps one atomic layer high \cite{Rousset2003VicinalSurfaces}.
Two recipes produced the aligned samples: low-coverage yielded one ribbon row per terrace,
whereas high-coverage yielded three rows per terrace
\cite{Darawish2025PrecursorCoverage}. High-coverage films on
200-nm Au(111)/mica chips were synthesised using the established high-throughput protocol
\cite{BorinBarin2019SwitchingDevices}. Section~S9 of the Supplementary Information
defines the backbone-to-backbone pitch, edge-to-edge distance, and steric-contact estimate.

\subsection{Transfer to the Raman-optimised substrate}

The Raman-optimised (RO) substrate comprises 40~nm of ALD-grown Al$_2$O$_3$ above an
approximately 80-nm metal layer on SiO$_2$/Si. Its optical design follows
Ref.~\cite{Overbeck2019OptimizedSubstrates}.

\paragraph{Polymer-free transfer from Au(111)/mica}
The \emph{Unaligned Au, high-coverage} film was transferred by the polymer-free float-off method
\cite{BorinBarin2019SwitchingDevices}. The chip was floated on aqueous HCl until
the mica delaminated. After dilution and rinsing, the floating GNR/Au film was collected with
the RO substrate, wetted with ethanol, annealed near $100\,^{\circ}$C for approximately 10~min,
and exposed to KI/I$_2$ to remove Au. This route produced configuration IV.

\paragraph{Electrochemical delamination from Au(788)}
The \emph{Aligned Au, high-coverage} film was transferred with a temporary PMMA support by
electrochemical bubble delamination \cite{Overbeck2019OptimizedSubstrates}.
The PMMA/GNR/Au sample served as the working electrode in 1~M NaOH. An applied potential of
approximately 5~V generated H$_2$ at the Au/electrolyte interface and detached the PMMA/GNR
film. The film was placed on the RO substrate, annealed sequentially near 80 and
$110\,^{\circ}$C, and rinsed after PMMA dissolution in acetone. This route produced
configuration V and retained the uniaxial array alignment.

\subsection{Temperature-dependent Raman spectroscopy (70--300~K)}

Raman spectra were acquired in backscattering geometry with a WITec Alpha 300 R confocal
microscope and a 785-nm continuous-wave laser. A $50\times$ long-working-distance objective
(NA $=0.55$, working distance 9.1~mm) produced an approximately $1.7~\mu$m spot. The incident
power density at the sample was $0.105~\mu\mathrm{W}\,\mu\mathrm{m}^{-2}$
($10.5~\mathrm{W}\,\mathrm{cm}^{-2}$). A 400-mm spectrometer with a
300-g\,mm$^{-1}$ grating and a cooled deep-depletion CCD recorded the spectra. The Si line at
$520.5~\mathrm{cm}^{-1}$ was used as a wavenumber reference before and after each series.
Calibration offset, dispersion accuracy and drift are not included in the
reported fitting uncertainties.

Measurements were performed in an Oxford Instruments Mercury iTC cryostat at
$4\times10^{-7}$~mbar. Acquisition started after at least 40~min of equilibration and after
the controller remained within $\pm0.05$~K for at least 5~min. A $z$ scan was used to select
the focal position giving the maximum $G$-band intensity, and the laser power was kept fixed
within each measurement series. Spatial sampling and integration settings varied between
configurations; the corresponding acquisition parameters are given in the Supplementary
Information. The controller stability describes the acquisition criterion; the uncertainty
of the ribbon temperature was not independently measured.

Repeated measurements were acquired from the same pre-registered sample region.
The complete Heating~1--Cooling~1--Heating~2 cycle was acquired only for
configuration II. The other configurations were measured along the heating path
identified in Table~S1. At each temperature, the spectral data used for analysis were
subjected to isolated-spike removal and asymmetric least-squares baseline subtraction.
Where multiple spatial spectra were available, the retained spectra were averaged
intensity-wise on their common Raman-shift grid; single-point measurements were analysed
from the accumulated spectrum. The resulting temperature-representative spectrum was
fitted globally, with no fit parameters shared between temperatures or configurations.
The preprocessing and global-fitting procedures are detailed in Sections~S1--S3 of the
Supplementary Information, and the peak-position, height, and linewidth definitions are
given in Section~S4.

\subsection{Spectral preprocessing and Lorentzian peak fitting}

Isolated spectral spikes were removed before background estimation, without broad smoothing
of the Raman bands (Fig.~S1). The background was then estimated by asymmetric
least squares (ALS) \cite{Eilers2005BaselineCorrection}. We used $p=0.0055$ and 10 iterations for every spectrum,
$\lambda=10^6$ for spectra measured on Au, and $\lambda=6\times10^3$ for both RO
configurations. The baseline was subtracted before fitting (Fig.~S2).

Peak parameters were obtained from full-window global fits to the baseline- and spike-corrected
spectrum at each temperature. For each temperature and array configuration, we fitted the full $200$--$2000~\mathrm{cm}^{-1}$
spectral window with a sum of Lorentzian profiles, one per Raman feature, as detailed in
Section~S3 of the Supplementary Information. The extracted peak position, peak height, and linewidth
from these single Lorentzian components (Section~S4) are used in all subsequent
temperature-dependent analyses. The fitted intensity was
\begin{equation}
 I_{\mathrm{fit}}(\omega)=\sum_{j=1}^{N}
 \frac{H_j}{1+4[(\omega-\omega_j)/\Gamma_j]^2},
 \label{eq:lorentzian_sum}
\end{equation}
where $N$ is the number of fitted components and $H_j$, $\omega_j$, and $\Gamma_j$ are the
height, centre frequency, and Lorentzian FWHM.
In the full-window fit, all components within a spectrum were optimised
simultaneously, and no fit parameter was shared between temperatures or
configurations. For configuration III,
the RBLM was subsequently refined locally, as described in Section~S3 of the Supplementary
Information. Additional components represented resolved structure in the
$1210$--$1355~\mathrm{cm}^{-1}$ interval; no physical assignment is made here.

Differential evolution \cite{Storn1997DifferentialEvolution} used a population
multiplier of 200, tolerance $10^{-7}$, and at most $10^5$ generations. Bounded nonlinear least
squares refined the global-search result. To prevent the numerous near-baseline points from
dominating the optimisation, residuals within predefined peak-profile windows received enhanced,
intensity-dependent weighting, while points elsewhere retained unit weight. The exact weighting
procedure is given in Section~S3 of the Supplementary Information. Regularisation included required-peak visibility penalties and additional
constraints on degenerate decompositions in the overlapping
$1211$--$1271~\mathrm{cm}^{-1}$ interval.
Reported $R^2$ and root-mean-square error (RMSE) values use unweighted full-window residuals. Parameter uncertainties were taken as the sample standard deviations
of 100 successful residual-bootstrap refits initiated from parameters perturbed by 3\% relative
to the optimum. The uncertainty procedure used for the configuration-III RBLM refinement is
described in Section~S3 of the Supplementary Information. Figures~S3--S5 show the fit bounds,
global decomposition, and Lorentzian metrics; Fig.~S7 shows $G$-window views and residuals of
the global fits for I and II at 80 and 300~K, selected to compare low- and high-coverage near
the ends of the measured interval.

\subsection{Temperature-dependent analysis and modelling}
\label{subsec:model}

For each temperature, the plotted peak position and linewidth are obtained from the accepted spectral fit
to the processed spectrum at that temperature. Their vertical bars are the one-standard-deviation
residual-bootstrap uncertainties of that fit. Frequency-model fits are unweighted, whereas the
linewidth fits use these uncertainties as relative weights.

The extrapolated intercept $\omega_0$ was fixed before fitting. For $D$ and $G$, an unweighted
linear fit to $T\leq110$~K was extrapolated to 0~K; for the RBLM, the unweighted linear fit used
all retained temperatures. For every mode, the reported $\omega_0$ uncertainty is the larger of
the propagated peak-position uncertainty and the intercept uncertainty obtained from
extrapolation residuals.

The temperature-dependent Raman peak position is modelled as
\begin{equation}
 \omega(T)=\omega_0+\Delta\omega_{\mathrm{TE}}(T)
 +\Delta\omega_{\mathrm{anh}}(T),
 \label{eq:omegaT}
\end{equation}
where $\omega_0$ is the precomputed extrapolated intercept defined above. The mismatch integral is
\begin{equation}
 \mathcal{I}(T)=\int_0^T
 [\alpha_{\mathrm{sub}}(T')-\alpha_{\mathrm{GNR}}(T')]\,dT',
 \label{eq:eps_mismatch}
\end{equation}
where $\alpha=(1/L)(dL/dT)$ is the linear thermal-expansion
coefficient (TEC). The mismatch therefore compares thermal expansion
along the nanoribbon axis with the in-plane thermal expansion of the
substrate.
Thus, $\mathcal{I}(T)$ is the accumulated difference between the free thermal expansions of the
substrate and ribbon from 0 to $T$. Assuming homogeneous, no-slip coupling between the ribbon and substrate,
the integrated thermal-expansion mismatch is taken to equal the axial strain transferred to the ribbon. The resulting thermoelastic shift of the Raman peak position is modelled as
\begin{equation}
 \Delta\omega_{\mathrm{TE}}(T)=
 \omega_0\{\exp[-\gamma_{\parallel}\mathcal{I}(T)]-1\}.
 \label{eq:Delta_TE}
\end{equation}
Here $\gamma_{\parallel}$ is an effective coupling parameter that converts the assumed axial
mismatch strain into a mode-frequency change; it is fitted separately for each mode and thermal
path.
For $|\gamma_{\parallel}\mathcal{I}|\ll1$,
\begin{equation}
 \Delta\omega_{\mathrm{TE}}(T)\simeq
 -\gamma_{\parallel}\omega_0\mathcal{I}(T).
 \label{eq:Delta_TE_linear}
\end{equation}
The exponential finite-strain expression reduces to the standard linear Gr\"uneisen--strain
relation in Eq.~\eqref{eq:Delta_TE_linear}
\cite{Yoon2011NegativeThermalExpansion,Bonini2007PhononAnharmonicities}.
To represent the temperature dependence associated with a lowest-order symmetric three-phonon
process, we use the phenomenological Klemens-type term
\begin{equation}
 \Delta\omega_{\mathrm{anh}}(T)=
 A_3\left[1+2n_{\mathrm{B}}\!\left(\frac{\omega_0}{2},T\right)\right],
 \label{eq:Delta_anh}
\end{equation}
with
\begin{equation}
 n_{\mathrm{B}}(\widetilde{\nu},T)=
 \left[\exp\!\left(\frac{hc\widetilde{\nu}}{k_{\mathrm{B}}T}\right)-1\right]^{-1}.
 \label{eq:BE}
\end{equation}
Here $\omega_0$ is the extrapolated reference Raman wavenumber of the
parent RBLM, $D$, or $G$ mode, whereas $\widetilde{\nu}$ denotes the
daughter-phonon wavenumber entering the Bose--Einstein occupation. In the
symmetric three-phonon Klemens process used in
Eq.~\eqref{eq:Delta_anh}, each daughter phonon has half the parent-mode
energy, so $\widetilde{\nu}=\omega_0/2$. Both wavenumbers are expressed in
$\mathrm{cm}^{-1}$, $h$ is the Planck constant, $k_{\mathrm{B}}$ is the
Boltzmann constant, and $c$ is expressed in
$\mathrm{cm}\,\mathrm{s}^{-1}$. Because $\omega_0$ is an empirical
extrapolated intercept rather than an independently calculated harmonic
frequency, $A_3$ is an effective coefficient: it includes a constant
intercept correction and is not interpreted as a uniquely isolated cubic
self-energy. Since $\Delta\omega_{\mathrm{anh}}(0)=A_3$, the unreferenced
model satisfies $\omega(0)=\omega_0+A_3$. Only the temperature-dependent
change in this fitted term, labelled \emph{anharmonic} in
Fig.~\ref{fig:Trends_Anh_TEC_Fits}, is used in the component decomposition.
It is referenced to the full model value at 0~K:
\begin{equation}
 \Delta\omega_{\mathrm{anh}}^{(0)}(T)=
 \Delta\omega_{\mathrm{anh}}(T)
 -\Delta\omega_{\mathrm{anh}}(0)
 =2A_3n_{\mathrm{B}}\!\left(\frac{\omega_0}{2},T\right).
 \label{eq:Delta_anh_referenced}
\end{equation}
Because $n_{\mathrm{B}}$ increases with temperature, $A_3<0$ makes this zero-K-referenced term
soften the mode in this sign convention, whereas $A_3>0$ makes it harden the mode.
The total zero-K-referenced change is consequently
\begin{equation}
 \Delta\omega^{(0)}(T)
 \equiv\omega(T)-\omega(0)
 =\Delta\omega_{\mathrm{TE}}(T)+\Delta\omega_{\mathrm{anh}}^{(0)}(T).
 \label{eq:Delta_total_referenced}
\end{equation}
Equations~\eqref{eq:Delta_anh} and~\eqref{eq:BE} follow the lowest-order Klemens functional form
\cite{Klemens1966AnharmonicDecay,Klemens1975Diamond}.

Au coefficients were obtained from the derivative of a fit to 22 literature
lattice parameters; the value at 300~K is
$\alpha_{\mathrm{Au}}=1.4668\times10^{-5}~\mathrm{K}^{-1}$
\cite{Pamato2018GoldExpansion}. 
We use the bulk sapphire value $\alpha_{\mathrm{Al_2O_3}}(300~\mathrm{K})=6.1724\times10^{-6}~\mathrm{K}^{-1}$ to approximate the in-plane thermal expansion of the Al$_2$O$_3$ surface layer
of the RO substrate
\cite{NISTSapphireCryogenic}.
Because the longitudinal TEC of 9-AGNRs is unavailable, the calculated axial
single-wall CNT coefficient, $\alpha_{\mathrm{CNT}}(300~K)=1.7837\times10^{-6}~\mathrm{K}^{-1}$, is used as a surrogate \cite{Jiang2009ThermalExpansion}. 

Figure~S6 documents the source data, adopted curves, and their mathematical low-temperature continuations. Above the 300~K endpoint
of the TEC grid, each coefficient is held at its 300~K value. At 300~K, the adopted differences
$\alpha_{\mathrm{Au}}-\alpha_{\mathrm{CNT}}=1.2884\times10^{-5}~\mathrm{K}^{-1}$ and
$\alpha_{\mathrm{Al_2O_3}}-\alpha_{\mathrm{CNT}}=4.39\times10^{-6}~\mathrm{K}^{-1}$;
the instantaneous mismatch coefficient is therefore 2.94 times larger for arrays on Au than for
the sapphire approximation to the Al$_2$O$_3$ layer of the RO substrate. The fit parameters are $A_3$ and
the effective coupling parameter $\gamma_{\parallel}$. Their uncertainties are one-standard-error estimates from the covariance matrix
of the unweighted fit, with $\omega_0$ and the TEC curves held fixed.
Uncertainties in the reference frequency, TEC inputs and sapphire/CNT
approximations are not propagated into these parameter errors (Section~S10).

For a compact comparison of the temperature response, Table~S2 reports two linear slopes in $\mathrm{cm}^{-1}\mathrm{K}^{-1}$. The data slope, $b_{\mathrm{data}}$, is obtained by fitting a straight line directly through the measured Raman peak positions. To obtain $b_{\mathrm{model}}$, the thermomechanical curve fitted using Eq.~\eqref{eq:omegaT} is evaluated at the same experimental temperatures, and a straight line is fitted through those values. Thus, $b_{\mathrm{data}}$ summarises the measured trend, whereas $b_{\mathrm{model}}$ summarises the trend described by the fitted thermomechanical model. Both quantities represent a single linear slope over the complete measured temperature range.

Linewidths were fitted with
\begin{equation}
 \Gamma_j(T)=\Gamma_{0,j}+A_{\Gamma,j}
 \left[1+2n_{\mathrm{B}}\!\left(
 \frac{\widetilde{\nu}_{j,\mathrm{LT}}}{2},T\right)\right],
 \label{eq:Gamma_anh}
\end{equation}
where $\widetilde{\nu}_{j,\mathrm{LT}}$ is the low-temperature reference
wavenumber of the parent Raman mode $j$. For each configuration and thermal path, it was fixed to the median fitted peak centre at the lowest measured temperature available for that series. In the symmetric three-phonon Klemens channel, each daughter phonon is assigned half the parent-mode energy; the Bose--Einstein occupation is therefore evaluated at
$\widetilde{\nu}_{j,\mathrm{LT}}/2$. The reference wavenumber was fixed
because its measured temperature variation is small relative to its
absolute value and the linewidth data do not independently constrain it.
Thus, only $\Gamma_{0,j}$ and $A_{\Gamma,j}$ were fitted.

Here, $\Gamma_{0,j}$ represents the temperature-independent residual
contribution, including static and instrumental broadening. Because the
retained three-phonon term includes its zero-point contribution,
$\Gamma_j(0)=\Gamma_{0,j}+A_{\Gamma,j}$. Only this lowest-order
three-phonon channel was retained. Equation~\eqref{eq:Gamma_anh} uses the
same Klemens population factor as Eq.~\eqref{eq:Delta_anh}
\cite{Klemens1966AnharmonicDecay,Klemens1975Diamond}.

\section{Results and Discussion}

\subsection{Sample configurations and low-temperature Raman fingerprints}

Figure~\ref{fig:Sample_Matrix} defines the five configurations used throughout this work:
(I) \emph{Aligned Au, low-coverage}; (II) \emph{Aligned Au, high-coverage};
(III) \emph{Unaligned Au, high-coverage}; (IV) \emph{Unaligned RO, high-coverage}; and
(V) \emph{Aligned RO, high-coverage}. Configuration V retained the principal
RBLM, $D$, and $G$ bands after transfer, but its $D$- and $G$-mode lines were
broader and its intermediate-frequency structure was less readily decomposed
than in its pre-transfer reference, II. The quantitative spectroscopic assessment is provided in Section~S14 of
the Supplementary Information. Accordingly, V is classified operationally as
a spectroscopically non-equivalent post-transfer sample; this designation
describes its changed Raman line shape and does not measure transfer yield,
efficiency, or atomic damage. V is therefore
retained as a diagnostic dataset but is not used in the principal mechanistic
comparisons, which concern I--IV. Figures~S1--S4 document representative
examples of spike removal, baseline subtraction, fitting windows, and global
Lorentzian decompositions for all five configurations, including V.

At the temperatures used in Fig.~\ref{fig:Sample_Matrix} (95, 100, or 105~K), all five
configurations display the RBLM, $D$, and $G$ features
(Fig.~\ref{fig:Sample_Matrix}). The RBLM lies between 309 and $314~\mathrm{cm}^{-1}$,
the $D$ mode between 1333 and $1345~\mathrm{cm}^{-1}$, and the $G$ mode between 1594 and
$1601~\mathrm{cm}^{-1}$. Additional intermediate-frequency features occur in the
$1210$--$1355~\mathrm{cm}^{-1}$ interval. Both transferred configurations retain
the RBLM, $D$, and $G$ features observed before transfer in their respective
source samples, III for IV and II for V.

\begin{figure}[!tp]
  \centering
  \includegraphics[width=\columnwidth]{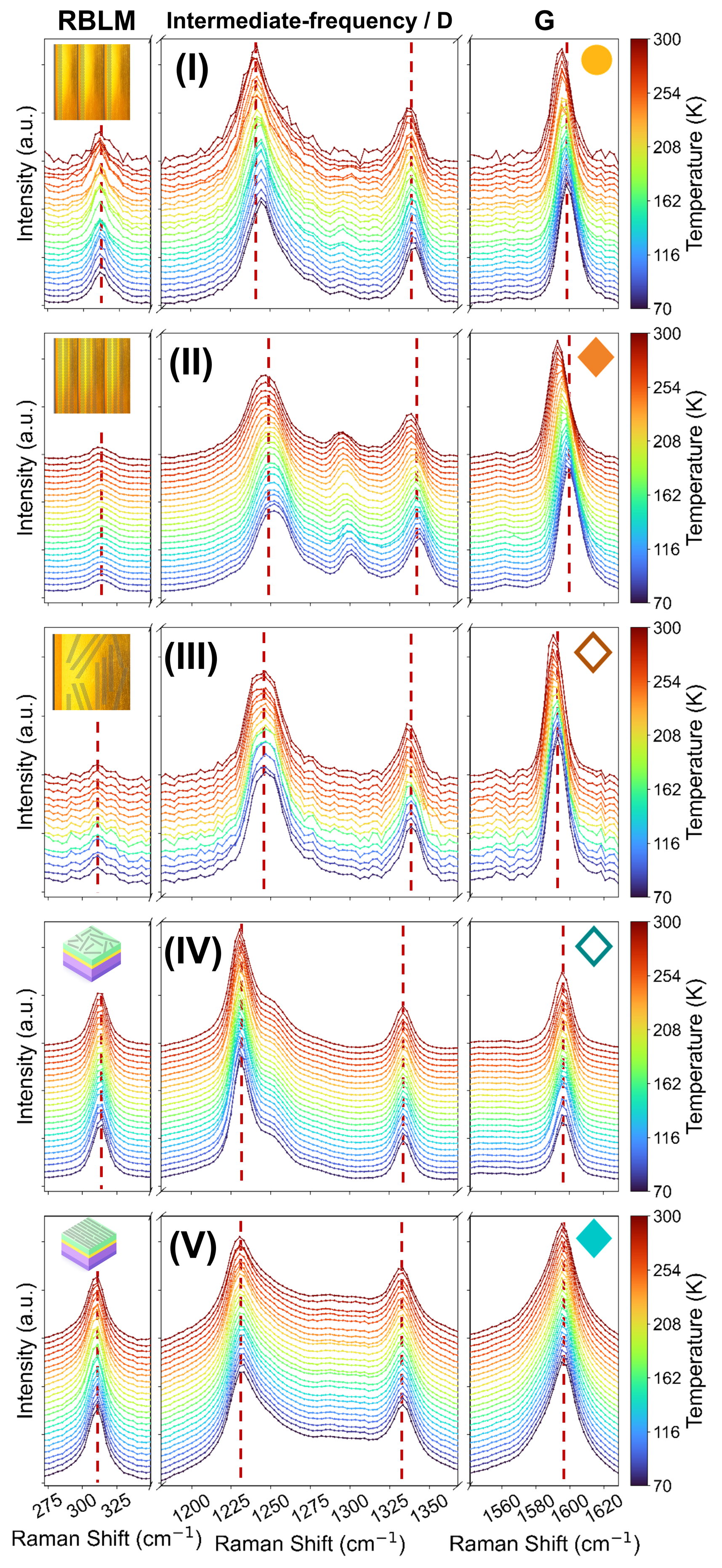}
  \caption{%
  Temperature-dependent, spike-removed and baseline-corrected Raman spectra
  ($\lambda=785$~nm) for (I) \emph{Aligned Au, low-coverage};
  (II) \emph{Aligned Au, high-coverage};
  (III) \emph{Unaligned Au, high-coverage};
  (IV) \emph{Unaligned RO, high-coverage}; and
  (V) \emph{Aligned RO, high-coverage}. Each row displays the RBLM,
combined intermediate-frequency/$D$, and $G$ spectral regions. Each spectrum is
  normalised to its own maximum and vertically offset. Colour denotes the
  actual measured temperature on a common scale spanning 70--300~K.
  Only measured spectra are shown. Fourteen spectra of V are resampled along the
  Raman-shift axis onto a common grid for display; no temperature measurement is
  interpolated, and the original grids are used for fitting.}
  \label{fig:Waterfall}
\end{figure}

Figure~\ref{fig:Waterfall} displays the measured spectral evolution for
configurations I--V. Table~S2 compares the measured temperature trend with
the trend described by the fitted thermomechanical model. For each Raman
mode, sample configuration, and reported thermal path, the data slope
$b_{\mathrm{data}}$ is obtained by fitting a straight line directly to the
measured peak positions. To obtain the model slope $b_{\mathrm{model}}$, the
fitted thermomechanical curve is first evaluated at the same experimental
temperatures, and a straight line is then fitted to those model values.
Thus, $b_{\mathrm{data}}$ summarises the average trend observed in the
measurements, whereas $b_{\mathrm{model}}$ summarises the average trend
described by the fitted model. Both quantities represent the average change
in Raman peak position per kelvin over the measured temperature range.

A negative slope denotes softening on heating. The straight-line $D$- and
$G$-mode slopes are negative for every main series, with smaller magnitudes
on the RO support (Table~S2). The modelled RBLM changes between 80 and 290~K
are below $1~\mathrm{cm}^{-1}$ (Table~S3), and several RBLM fits have lower
$R^2$ values and larger relative parameter errors (Table~S1).

\subsection{Temperature-dependent phonon frequencies}

\begin{figure}[!tp]
  \centering
  \includegraphics[width=\columnwidth]{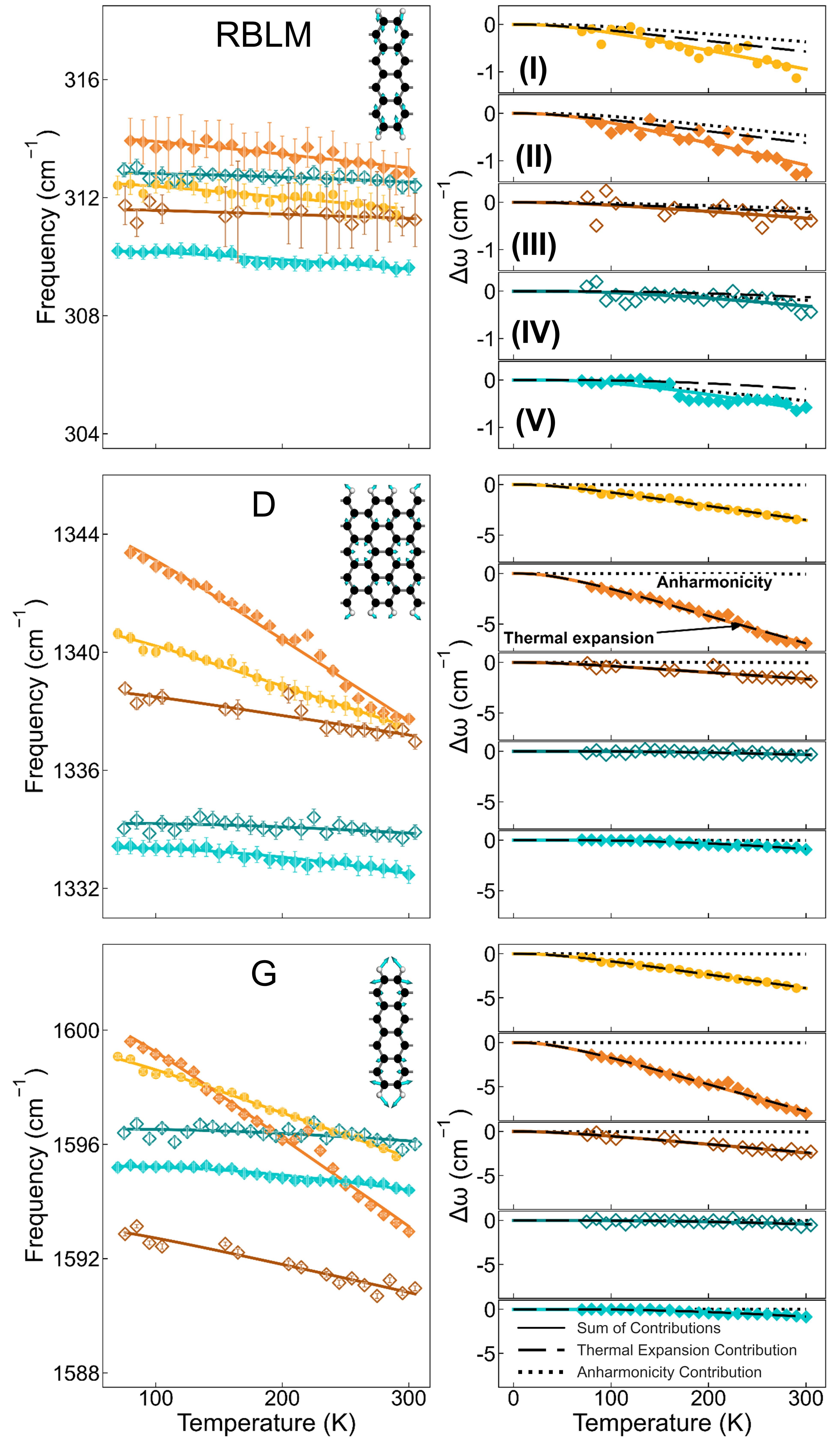}
  \caption{%
  Temperature dependence of the Lorentzian peak positions of the RBLM
  (top), $D$ feature (middle), and $G$ mode (bottom) for configurations
  I--V. Insets schematically illustrate representative atomic displacement
  patterns associated with the three Raman features; cyan arrows indicate
  the displacement directions. Colours and symbols follow
  Fig.~\ref{fig:Sample_Matrix}. The five stacked right-hand panels in each
  row are ordered I--V from top to bottom. Left: measured Raman peak
  positions $\omega(T)$ and fits to Eq.~\eqref{eq:omegaT}. Vertical bars
  show one-standard-deviation residual-bootstrap uncertainties; the fits
  are unweighted. Right: absolute Raman peak-position changes
  $\omega(T)-\omega(0)$ in cm$^{-1}$, where
  $\omega(0)=\omega_0+A_3$. Coloured solid curves show the total fitted
  response, black long-dashed curves the \emph{thermoelastic}
  contribution, and black dotted curves the zero-K-referenced
  Klemens-type contribution, labelled \emph{anharmonic} in the panels.}
  \label{fig:Trends_Anh_TEC_Fits}
\end{figure}

Figure~\ref{fig:Trends_Anh_TEC_Fits} shows the temperature dependence of the
three Raman peak positions. The most direct transfer comparison is
III$\rightarrow$IV, for which the same unaligned, high-coverage film was
measured on Au and again after polymer-free transfer to the RO substrate.
The straight-line slopes through the measured peak positions are
$-0.0064$ and $-0.0014~\mathrm{cm}^{-1}\mathrm{K}^{-1}$ for $D$ in III
and IV, and $-0.0094$ and $-0.0017~\mathrm{cm}^{-1}\mathrm{K}^{-1}$ for
$G$, respectively (Table~S2). Both modes therefore have smaller redshift rates
after transfer of the same film to the RO support. The principal Raman bands
persist after transfer, retaining the characteristic Raman fingerprint.
Changes in substrate contact, residual strain and strain transfer can contribute
to the altered temperature response.

Configurations II and III provide a complementary comparison between two
high-coverage 9-AGNR arrays on Au. The measured $D$- and $G$-mode
redshift rates are larger in II (Table~S2).
Within the adopted model, both configurations use the same Au and ribbon TEC
curves, so their different responses are absorbed into the fitted effective coupling
coefficients. This is a model-based description; the data alone do not determine
whether the difference arises from constraint, adsorption, residual strain, or
the distinct bulk-Au and Au/mica supporting structures.

Previous Raman and STM measurements of 9-AGNRs on Au(788) suggested that
ribbons occupying the lower step edges interact more strongly with Au,
which was attributed to the greater reactivity of the lower-step Au sites
\cite{Darawish2024AlignmentQuality}. The periodic
step-edge adsorption sites in II therefore provide a physical basis for
stronger ribbon--substrate constraint and more effective transfer of the
thermally generated mismatch strain. Configuration III lacks this regular
vicinal step template. Because alignment, step-edge pinning, adsorption
registry, and substrate architecture change together between II and III,
the comparison is consistent with configuration-dependent strain transfer, but
neither measures it independently nor isolates ribbon alignment as its sole origin.

\begin{figure}[!t]
  \centering
  \includegraphics[width=\columnwidth]{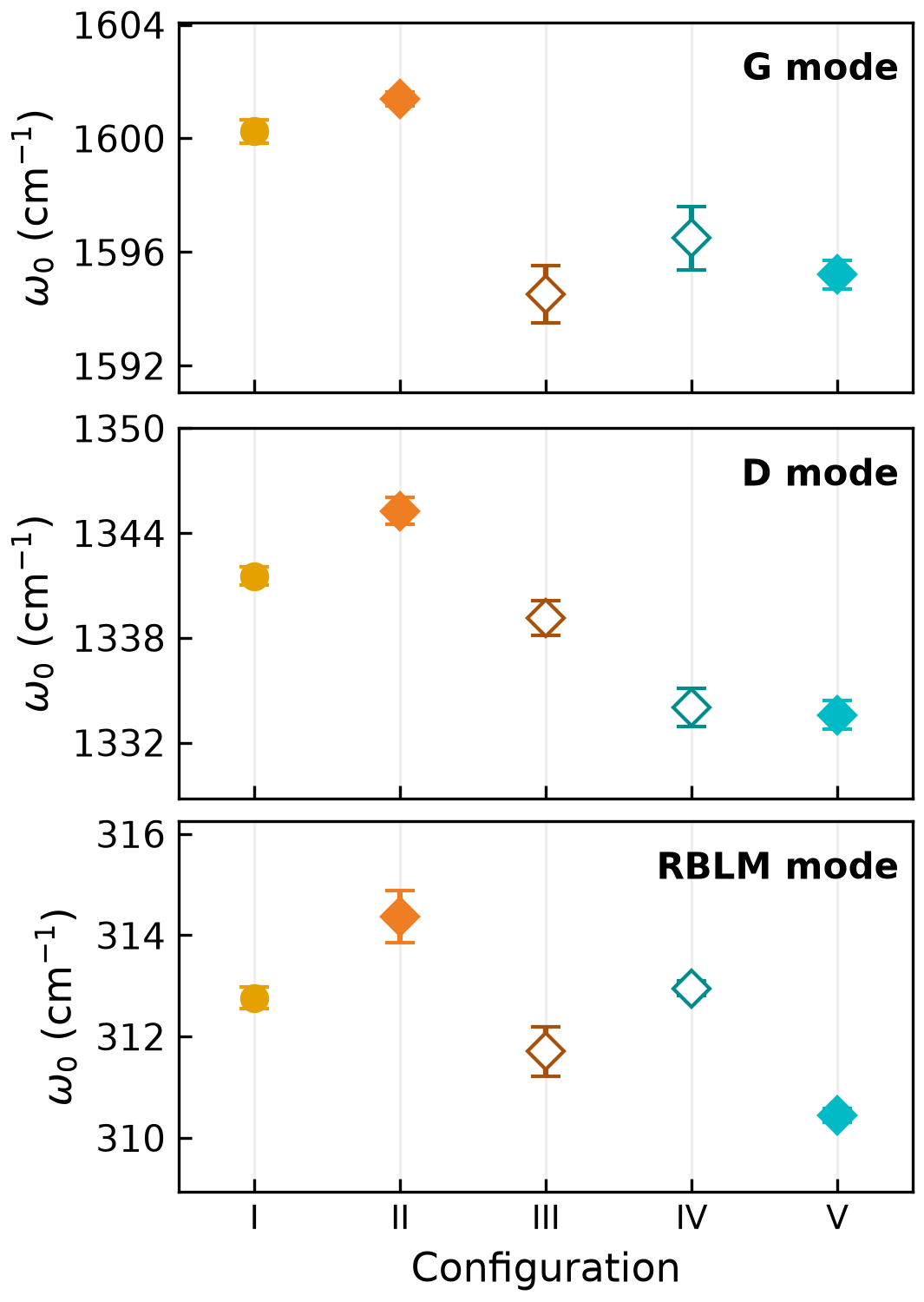}
  \caption{%
  Extrapolated reference frequencies $\omega_0$ for the $G$, $D$, and RBLM
  modes of I--V. For $D$ and $G$, each point is the intercept of an
  unweighted linear fit to measurements with $T\leq110$~K. Because the RBLM
  positions exhibit greater scatter, all retained temperatures are used to
  obtain a stable RBLM intercept. Vertical bars are one-standard-error
  intercept uncertainties. For each series, the larger of (a) the
  uncertainty obtained by propagating the measured peak-position
  uncertainties and (b) the uncertainty obtained from the extrapolation
  residuals is shown.}
  \label{fig:omega0_comparison}
\end{figure}

Whereas Fig.~\ref{fig:Trends_Anh_TEC_Fits} quantifies the temperature
response, Fig.~\ref{fig:omega0_comparison} compares the extrapolated
reference frequencies $\omega_0$ used in the temperature model; Table~S1
lists their values. The quantity $\omega_0$ is an empirical extrapolated
reference.

Configurations I (\emph{Aligned Au, low-coverage}) and II
(\emph{Aligned Au, high-coverage}) provide the most direct comparison of
coverage and lateral packing. The ribbon width, edge termination, Au(788)
substrate, and step-directed alignment are retained, while the precursor
dose changes the number of ribbon rows occupying each terrace. This packing
contrast is directly visible in the STM images and corresponding schematics
in Fig.~\ref{fig:Sample_Matrix}(a), and agrees with the reported
precursor-dose-dependent growth of 9-AGNRs on Au(788)
\cite{Darawish2025PrecursorCoverage}. At low precursor dose, the
ribbons preferentially form one row along the lower step edge of each
terrace, whereas a complete high-coverage monolayer contains three parallel
rows per terrace. The terrace occupancy establishes a denser array. Dividing a terrace width
of $3.8$--$3.9$~nm by three gives an average row period of about $1.3$~nm.
Local separations can vary with registry and height. Section~S9 gives a separate
illustrative H-contact estimate under explicit geometric assumptions.

All three extrapolated reference-frequency estimates are higher in II than
in I: $1.60\pm0.55~\mathrm{cm}^{-1}$ for RBLM,
$3.74\pm0.92~\mathrm{cm}^{-1}$ for $D$, and
$1.14\pm0.49~\mathrm{cm}^{-1}$ for $G$. These errors combine the reference
uncertainties defined in Section~S10 in quadrature. The smaller $G$ offset
has a larger relative uncertainty. Its measured frequencies cross between
130 and 140~K, whereas the RBLM remains higher in II at the common measured
temperatures.

Within a harmonic description,
$\omega_j\propto\sqrt{k_{\mathrm{eff},j}/\mu_j}$, where
$k_{\mathrm{eff},j}$ is the effective restoring force associated with mode
$j$ and $\mu_j$ is its participating mass
\cite{Wilson1955MolecularVibrations}. Because I and II contain
the same 9-AGNR structure and edge chemistry, their participating atomic
masses are effectively unchanged. For modes with comparable character, the
upward reference-frequency estimates are consistent with modest increases in the effective
modal restoring forces as the array becomes more densely packed. Here,
stiffening refers to the restoring force of the complete normal mode and
does not require a particular C--C bond to become shorter.

The RBLM probes a collective transverse width oscillation and is therefore
sensitive to the lateral environment. Repulsive interactions between sufficiently close H-terminated edges could
increase its effective restoring force. Calculations for interacting edges
within curved ribbons illustrate this possibility \cite{Chang2014EdgeDecorated};
their application to adjacent flat 9-AGNRs on Au remains qualitative. The $D$ and $G$
features probe edge-activated and backbone vibrations, respectively. Their
reference-frequency shifts may also include altered adsorption and residual strain.

Independent Raman evidence further supports the sensitivity of the
intermediate-frequency vibrational response of 9-AGNRs to lateral packing.
A previous study on Au(788) reported a feature near
$1241~\mathrm{cm}^{-1}$ at high-coverage and
$1235~\mathrm{cm}^{-1}$ at low-coverage, together with an additional feature
near $1285~\mathrm{cm}^{-1}$ observed only at high-coverage
\cite{Darawish2024AlignmentQuality}. Recent resonance Raman work on 7-AGNRs
assigns features in this region to intrinsic zone-folded phonons, providing
a width-specific comparison rather than a unique C--H assignment for 9-AGNRs
\cite{Nascimento2025PhononAssignment}. The observed
coverage-dependent changes show that this vibrational region is sensitive
to the lateral ribbon environment. Higher coverage can
also alter the occupation of terrace-centre and upper-step adsorption sites,
so adsorption registry and residual static strain may influence the exact
frequency shifts. The denser array and higher reference-frequency estimates are therefore
consistent with a coverage-dependent change in modal stiffness, while lateral
interactions are not isolated from adsorption and residual strain.

\subsection{Thermoelastic and anharmonic contributions to the Raman shifts}

Table~S3 compares the modelled changes between 80 and 290~K. For I--IV,
the $D/G$ changes range from about $-6.32~\mathrm{cm}^{-1}$ for $G$ in II
to $-0.30~\mathrm{cm}^{-1}$ for $D$ in IV. With the fixed-reference
convention, the model assigns most of the $D/G$ shift to the thermoelastic
term. This is consistent with a substantial contribution from thermal-expansion
mismatch. The fitted partition depends on the empirical reference and the
mechanical approximations used for the ribbon and support.

The adopted TEC curves give an instantaneous mismatch ratio of 2.94 at 300~K
and an accumulated 80--290~K ratio of 4.45. The corresponding accumulated
mismatches are $0.257\%$ on Au and $0.0578\%$ for the sapphire approximation
to the RO surface. These estimates illustrate the smaller mismatch on the RO
support. Quantitative strain transfer also depends on the interface and on the
mechanical constraint imposed by the underlying layers.

The main $D/G$ redshift persists through all three measured paths of II
(Table~S4). Differences between the paths describe one cycle of the same
sample; additional cycles and quantified drift would be needed to establish
reproducible thermodynamic hysteresis. Agreement between the data and model
slopes checks the consistency of the fitted description.

\subsection{Temperature dependence of Raman linewidths}

For a homogeneous Lorentzian line, a larger FWHM corresponds to a higher
phonon dephasing rate and a shorter coherence time. Only in the
lifetime-limited case, when pure dephasing and other broadening mechanisms
are negligible, can the linewidth be converted directly into a phonon
population lifetime. Anharmonic phonon decay nevertheless provides a
natural source of thermal broadening. Increasing temperature raises the
Bose--Einstein occupation of lower-energy daughter phonons, enhances
stimulated phonon--phonon decay and scattering, and broadens the homogeneous
line. The FWHM obtained from the present Lorentzian fits can also contain
instrumental, spatially inhomogeneous, and unresolved line-shape broadening.
Because these contributions are not separated, the fitted FWHM is treated
as an effective spectral linewidth and is not converted into a phonon
lifetime.

Figure~S9 reports the effective RBLM, $D$, and $G$ FWHMs for all five
configurations. In the dense aligned array II, the $G$ peak narrows during
the first part of Heating~1 and broadens again at higher temperatures.
Cooling and reheating show related non-monotonic trends, with different
depths and positions (Fig.~S10). The corresponding $D$ changes are smaller
relative to their fitting uncertainties. The first-heating $G$ result is
therefore the clearest observation of narrowing followed by broadening.
All three paths belong to one measured cycle of the same sample.

The cubic Klemens term in Eq.~\eqref{eq:Gamma_anh} contains the population
factor $1+2n_{\mathrm B}(\widetilde{\nu}_{j,\mathrm{LT}}/2,T)$, where
$n_{\mathrm B}$ is defined in Eq.~\eqref{eq:BE}. Because
$n_{\mathrm B}$ increases monotonically with temperature,
Eq.~\eqref{eq:Gamma_anh} is monotonic for any fixed
$A_{\Gamma,j}$. A positive cubic decay amplitude produces progressive
broadening, whereas a negative amplitude produces progressive narrowing.
Neither case can generate an intermediate-temperature minimum. A resolved
intermediate-temperature minimum therefore cannot be described by a single
cubic Klemens channel.

For comparison, the standard quartic, or four-phonon, contribution to the
linewidth of mode $j$ can be written as
\begin{equation}
 \Gamma_{4,j}(T)=A_{\Gamma,4,j}
 \left[
  1+3n_{\mathrm B}\!\left(
  \frac{\widetilde{\nu}_{j,\mathrm{LT}}}{3},T\right)
  +3n_{\mathrm B}^{\,2}\!\left(
  \frac{\widetilde{\nu}_{j,\mathrm{LT}}}{3},T\right)
 \right],
 \label{eq:Gamma_quartic}
\end{equation}
where $A_{\Gamma,4,j}$ is the quartic broadening amplitude of mode $j$. For
$A_{\Gamma,4,j}>0$, Eq.~\eqref{eq:Gamma_quartic} is also monotonic. In the
high-temperature limit, $n_{\mathrm B}\propto T$, so its quadratic
population term causes the quartic contribution to grow approximately as
$T^2$. Four-phonon broadening can therefore become increasingly important
and more readily distinguishable from the cubic contribution as the upper
temperature is extended
\cite{Balkanski1983AnharmonicEffects}.

Guo \emph{et al.} analysed 7-AGNRs on Au(111) over 80--520~K and included
cubic and quartic terms in the nonlinear RBLM frequency and linewidth
models, with the quartic contribution found to be dominant. Their higher
upper temperature provided greater sensitivity to the more rapidly growing
quartic term than is available in the present measurements, which end near
300~K. They also reported a turnover of the 7-AGNR $G$-mode linewidth near
180~K, reproduced it with an empirical second-order polynomial, and
qualitatively discussed the trend as competition between electron--phonon
and phonon--phonon contributions
\cite{Guo2022PhononAnharmonicities}. These observations provide an AGNR precedent for competing temperature-dependent
linewidth contributions. In the present range, the cubic term serves as the
anharmonic baseline. A positive quartic term would add broadening, so explaining
the initial narrowing requires a contribution that decreases with temperature
or a change in spectral line shape.

The electron--phonon interpretation proposed by Guo \emph{et al.} requires
a distinction between general electron--phonon coupling and direct phonon
decay into an electron--hole pair. Electron--phonon coupling arises when
the atomic displacements associated with a phonon perturb the electronic
potential and thereby couple electronic states
\cite{Lee2023EPW}. The real part of the
resulting phonon self-energy can shift the phonon frequency, whereas a
linewidth contribution from electron--hole-pair creation requires a real
electronic transition that satisfies the energy and momentum conditions.
Electron--phonon coupling can therefore remain present even when direct
electron--hole-pair creation is energetically forbidden.

In gapless graphite, low-energy electronic transitions are available and a
$G$ phonon can decay by creating an electron--hole pair. As temperature
increases, Fermi--Dirac broadening reduces the occupation difference between
the initial and final electronic states, causing this electron--hole-pair
damping contribution to decrease. Liu \emph{et al.} assigned the minimum in
the graphite $G$-mode FWHM near 700~K to competition between this decreasing
electronic contribution and increasing three- and four-phonon broadening
\cite{Liu2019GraphiteRaman}. This mechanism has a
microscopic basis in gapless graphite but cannot be assumed solely from the
presence of electron--phonon coupling in a semiconducting nanoribbon.

For the present 9-AGNRs, a $G$ phonon near
$1600~\mathrm{cm}^{-1}$ has an energy
\[
 hc\widetilde{\nu}_G\approx0.20~\mathrm{eV}.
\]
Scanning-tunnelling spectroscopy of Au(111)-supported 9-AGNRs reports an
electronic gap of
$1.4~\mathrm{eV}$
\cite{Talirz2017NineAGNR}. For electronic states retaining that gap, a single $G$ phonon is
energetically incapable of creating a valence-to-conduction electron--hole
pair. This Au(111) gap is not a measurement of every configuration;
substrate-hybridised states, doping, defects, and intraband channels are not
excluded by this energy comparison. In their 7-AGNR study,
Guo \emph{et al.} explicitly neglected electron--hole-pair creation when
analysing the RBLM linewidth because the ribbon gap is approximately
$2.3$~eV. The same energy restriction applies to a
$\sim0.20$~eV $G$ phonon in a wide-gap AGNR. Their empirical $G$-mode
turnover therefore remains a relevant precedent for competing linewidth
contributions, but their qualitative electron--phonon attribution does not
establish a graphite-like electron--hole-pair decay mechanism.

The first-heating $G$ narrowing thus points to an additional temperature-dependent
contribution beyond conventional anharmonic broadening. The measured linewidth
combines phonon damping, instrumental broadening, spatial variation and unresolved
spectral components. Determining their individual contributions remains an open
question.

\subsection{Overall discussion and model limitations}
\label{subsec:overall_discussion}

The matched III$\rightarrow$IV comparison shows that the same unaligned film
has smaller $D$- and $G$-mode redshift rates after transfer from Au to the RO
support. This substrate dependence is consistent with a substantial thermoelastic
contribution. The model describes it through an effective coupling coefficient
that combines mode strain sensitivity with interfacial constraint. The division
between thermoelastic and anharmonic contributions depends on the empirical
reference frequency and the expansion curves used for the ribbon and support.

Increasing coverage on aligned Au is accompanied by higher extrapolated reference
frequencies, most clearly for the RBLM and $D$ modes. For comparable modes, these
shifts are consistent with stronger effective restoring forces in the denser
array. Terrace occupancy, adsorption geometry and residual strain also change
with coverage, so the contribution of lateral ribbon interactions remains to be
separated from these effects. The smaller $G$ reference offset concerns the
low-temperature extrapolation; the measured frequencies cross as temperature rises.
Similarly, the aligned and unaligned Au arrays differ in their step geometry,
which prevents assigning their different responses to alignment alone.

The dense aligned array also shows a $G$ peak that narrows on first heating and
broadens again at higher temperatures. Conventional positive cubic and quartic
anharmonic terms increase with temperature, so the initial narrowing requires an
additional contribution or a change in spectral line shape. The corresponding
$D$ trend is weaker. Local strain variations, unresolved overlapping vibrations
and phonon damping can all affect the measured widths, and their individual
contributions remain to be determined.

The model uses bulk sapphire and an axial CNT as approximations for the layered
RO support and the ribbon. The reported errors describe fitting precision under
these assumptions; calibration drift, sample-temperature accuracy and variability
between films require further measurements. Configuration V, with broader $D/G$
lines after transfer, illustrates the sensitivity of Raman spectra to changes in
the sample and its interface. Together, the frequency and linewidth results show
why substrate choice and array arrangement matter when interpreting and tailoring
supported nanoribbon vibrations. Independently varying ribbon separation and
substrate constraint, together with spatially resolved Raman measurements,
would help distinguish these contributions.

\section{Conclusions}

Temperature-dependent Raman measurements show that the vibrational response of
supported 9-AGNR arrays depends on the substrate and ribbon coverage. The same
unaligned film has smaller $D/G$ redshift rates after transfer from Au to the RO
support, consistent with a substantial thermoelastic contribution at the interface.
Its separation from anharmonic effects remains model-dependent. The dense aligned
Au array has higher extrapolated reference frequencies, most clearly for the RBLM
and $D$ modes, and its $G$ peak narrows on first heating before broadening at higher
temperatures. This linewidth behaviour points to an additional contribution beyond
conventional anharmonic broadening; evidence for a corresponding $D$ minimum is
weaker. These results provide a basis for using substrate choice and ribbon packing
to influence nanoribbon properties after synthesis.

\section*{Data availability}
Supplementary Information accompanies this article. The research data
underlying the main text and Supplementary Information are available on
Zenodo at \url{https://doi.org/10.5281/zenodo.23040882}. Analysis code is
available at
\url{https://github.com/labordet/agnr-raman-phonons/releases/tag/v1.0-paper}.

\section*{Acknowledgements}
We warmly thank Amogh Kinikar for his extensive training of ALA on the GNR reactor and the Au/mica transfer process, and for his generous practical advice.

GBB and MC acknowledge financial support from the European Union's Horizon Europe research and innovation programme under Grant Agreement No.~101099098 (ATYPIQUAL). GBB also acknowledges financial support from the Werner Siemens Foundation (CarboQuant project), and the State Secretariat for Education, Research and Innovation (SERI) under Contract No.~23.00422.

\section*{Author contributions}
ALA, MC, and MD conceived and designed the experiments. 
GBB and ALA synthesised the 9-AGNR samples. 
GBB carried out the transfer of the high-coverage aligned films from Au(788) 
to the RO substrate [configuration II$\to$V], and ALA carried out the 
transfer of the unaligned films from Au(111) to the RO substrate 
[configuration III$\to$IV]. 
GBB acquired the STM images shown in Fig.~1(a); MC prepared the corresponding STM schematic illustrations, and ALA and MC prepared the schematic illustrations of the RO substrate. 
MD acquired the Raman data for configurations III and IV, while ALA acquired 
the Raman data for configurations I, II, and V. 
ALA performed all spectral preprocessing, global peak fitting, and 
thermomechanical and linewidth analysis. 
ALA curated and deposited the raw, intermediate, and processed data on Zenodo,
and developed and documented the Raman-processing and analysis code released through GitHub.
ALA prepared the remaining figures and wrote the manuscript with input from MC and MD. 
MC and MD supervised the project. 
All authors discussed the results and commented on the manuscript.

\appendix



\bibliographystyle{elsarticle-harv}
\bibliography{references}






\end{document}


\begin{frontmatter}
\title{Supplementary Information for\\
Surface-Dependent Phonon Dynamics in 9-Armchair Graphene Nanoribbon Arrays}
\end{frontmatter}

\section*{Contents of the Supplementary Information}

This Supplementary Information documents the Raman-spectrum treatment,
Lorentzian line-shape analysis, and temperature-dependent modelling. Sections~S1--S4
and Figs.~\ref{fig:spike_removal}--\ref{fig:lorentzian_metrics} describe isolated-spike
removal, asymmetric-least-squares background subtraction, the fitted peak-position
windows, representative full-window Lorentzian decompositions, and the extracted
Lorentzian parameters. Section~S5 and Fig.~\ref{fig:TEC} document the thermal-expansion
coefficients used in the thermomechanical model. Section~S6 and
Fig.~\ref{fig:G_residuals} examine representative $G$-region residuals. Sections~S7
and~S13, Fig.~\ref{fig:thermal_cycles}, and
Table~\ref{tab:thermal_path_comparison} report the path-resolved thermal-cycle analysis.
Section~S8 and Figs.~\ref{fig:linewidths} and~\ref{fig:linewidth_cycles} report the
linewidth analysis. Section~S9 distinguishes average row spacing from an illustrative edge-contact estimate.
Tables~\ref{tab:model_params}--\ref{tab:shift_decomposition} report the fitted
thermomechanical parameters, linear temperature slopes, and additive decomposition of
the modelled 80--290~K Raman peak-position shifts into thermoelastic and effective
Klemens-type contributions. Section~S14 provides the quantitative post-transfer
spectroscopic assessment of configuration V.

The Raman measurements were performed on five sample configurations distinguished by ribbon
alignment, coverage, and substrate: \emph{Aligned Au, low-coverage}; \emph{Aligned Au, high-coverage}; \emph{Unaligned Au, high-coverage}; \emph{Unaligned RO, high-coverage}; and
\emph{Aligned RO, high-coverage}. Here ``RO'' denotes the \RO. Temperature-dependent Raman
spectra were acquired at successive equilibrated temperature points along heating or cooling
paths. For the \emph{Aligned Au, high-coverage}
configuration, a complete heating, cooling, and reheating cycle was measured to evaluate
thermal-cycle reproducibility, as shown in Fig.~\ref{fig:thermal_cycles}.

\subsection*{Raman acquisition conditions}

All temperature-dependent measurements used 785-nm excitation, a
300-g\,mm$^{-1}$ grating, and the same $50\times$ long-working-distance
objective. Spatial sampling and integration differed between configurations,
as summarised below.

\begin{center}
\small
\renewcommand{\arraystretch}{1.15}
\begin{tabular}{p{0.07\textwidth} p{0.28\textwidth} p{0.53\textwidth}}
\toprule
\textbf{ID} & \textbf{Sample configuration} & \textbf{Acquisition} \\
\midrule

I &
Aligned Au, low-coverage &
2-$\mu$m line scan with 12 spatial points; 40~s integration per point. \\

II &
Aligned Au, high-coverage &
Single-point measurement; 14 accumulations of 60~s each. \\

III &
Unaligned Au, high-coverage &
Single-point measurement; 20 accumulations of 60~s each. \\

IV &
Unaligned RO, high-coverage &
69.73-$\mu$m line scan with 20 spatial points; 1~s integration per point. \\

V &
Aligned RO, high-coverage &
10$\times$10~$\mu$m$^2$ map sampled on a 5$\times$5 grid
(25 spatial points); 10~s integration per point. \\

\bottomrule
\end{tabular}
\end{center}

Absolute Raman intensities are not used for comparisons between configurations;
the temperature-dependent analysis is based on the fitted Raman peak positions
and linewidths. Spectra shown for comparison are normalised as stated in the
corresponding figure captions.

For configuration III, 20 of 28 candidate acquisitions were retained for
temperature averaging; the omitted acquisitions generally showed substantially
weaker Raman signal and were excluded before temperature-dependent fitting.
The complete inclusion mask is provided with the deposited research data.

\section{Removal of isolated spectral spikes}

Sharp, isolated spikes in intensity are removed before background estimation. These anomalous points can distort the estimated baseline and destabilise the subsequent nonlinear fit. The correction is restricted to the narrow affected region, and no broad smoothing is applied to the Raman bands. The measured Raman-shift grid is preserved. The local intensity corrections are documented in the processing records.

\begin{figure}[p]
  \centering
  \includegraphics[width=0.96\textwidth]{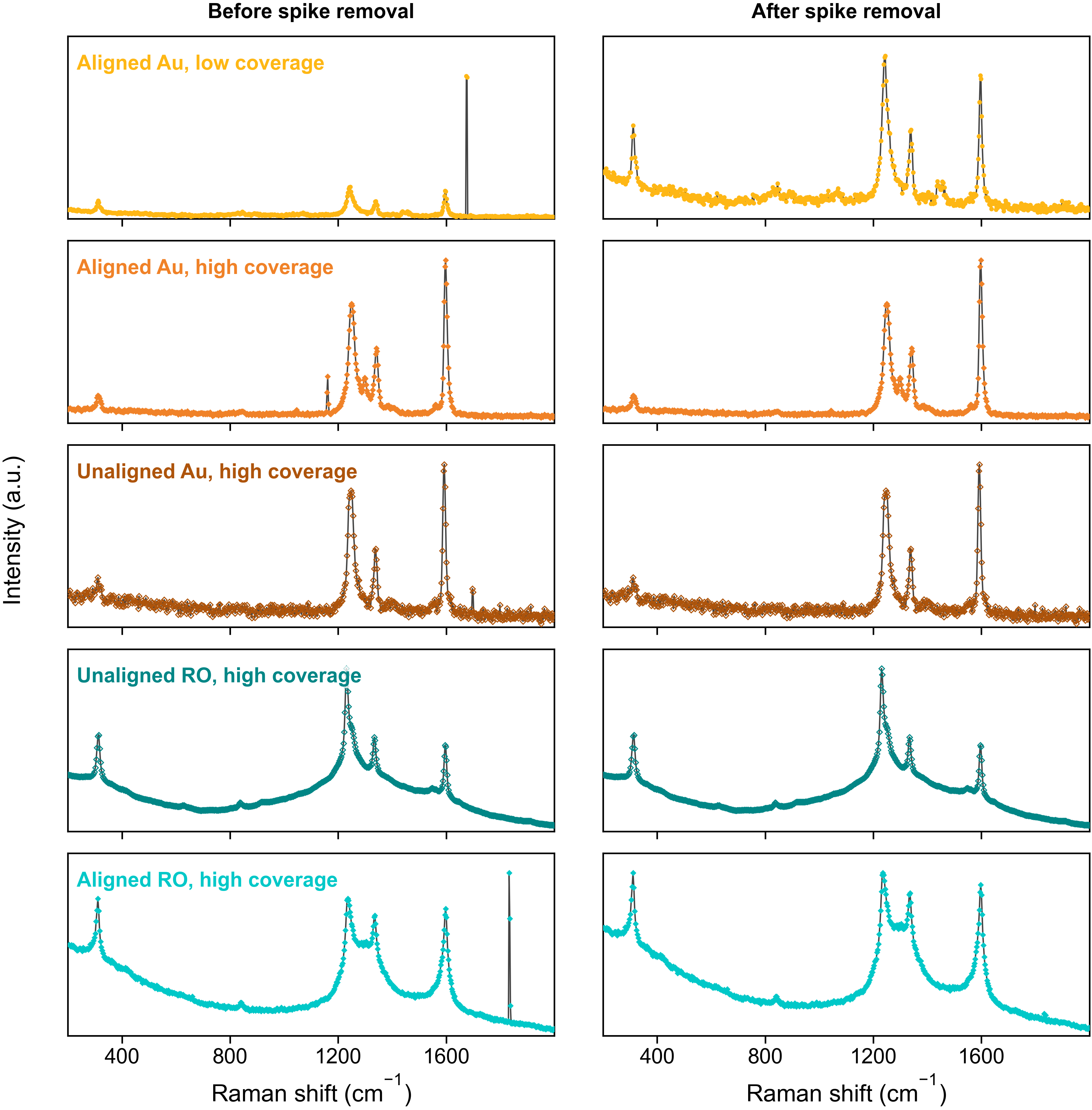}
  \caption{%
  Spike removal for representative Raman spectra from all five sample configurations. Left column:
  spectra before spike removal. Right column: the corresponding spectra after removal of
  isolated spectral spikes. From top to bottom the rows show \emph{Aligned Au, low-coverage};
  \emph{Aligned Au, high-coverage}; \emph{Unaligned Au, high-coverage};
  \emph{Unaligned RO, high-coverage}; and \emph{Aligned RO, high-coverage}. The operation
  suppresses narrow isolated spikes while preserving
  the RBLM, $D$-region, and $G$-mode features.}
  \label{fig:spike_removal}
\end{figure}

Figure~\ref{fig:spike_removal} compares representative spectra from all five sample configurations
before and after this correction. The isolated intensity spikes visible in the left-hand panels
are removed without broad smoothing of the Raman bands.

\FloatBarrier

\section{Baseline removal by asymmetric least squares}

After spike removal, the slowly varying spectral background is estimated by asymmetric least
squares (ALS). For measured intensities $y_i$ and baseline values $z_i$, ALS determines a smooth
baseline $z$ by minimising
\begin{equation}
  \sum_i w_i\left(y_i-z_i\right)^2
  + \lambda\sum_i\left(\Delta^2 z_i\right)^2,
  \label{eq:ALS}
\end{equation}
where $\lambda$ controls baseline smoothness. At each iteration, $w_i=p$ when $y_i>z_i$ and
$w_i=1-p$ otherwise. Positive Raman peaks therefore receive a smaller weight than points on or
below the estimated background and have reduced influence on the fitted baseline. The estimated baseline is
then subtracted from the spike-corrected spectrum before Lorentzian fitting.

For all sample configurations, ALS baseline correction was performed using an asymmetry parameter of $p = 0.0055$ and 10 iterations. A smoothness parameter of $\lambda = 10^{6}$ was used for all Au-supported samples, whereas $\lambda = 6 \times 10^{3}$ was used for both Raman-optimised samples. Subsequent analysis was performed over the $200$--$2000~\mathrm{cm}^{-1}$ spectral interval.

\begin{figure}[p]
  \centering
  \includegraphics[width=0.96\textwidth]{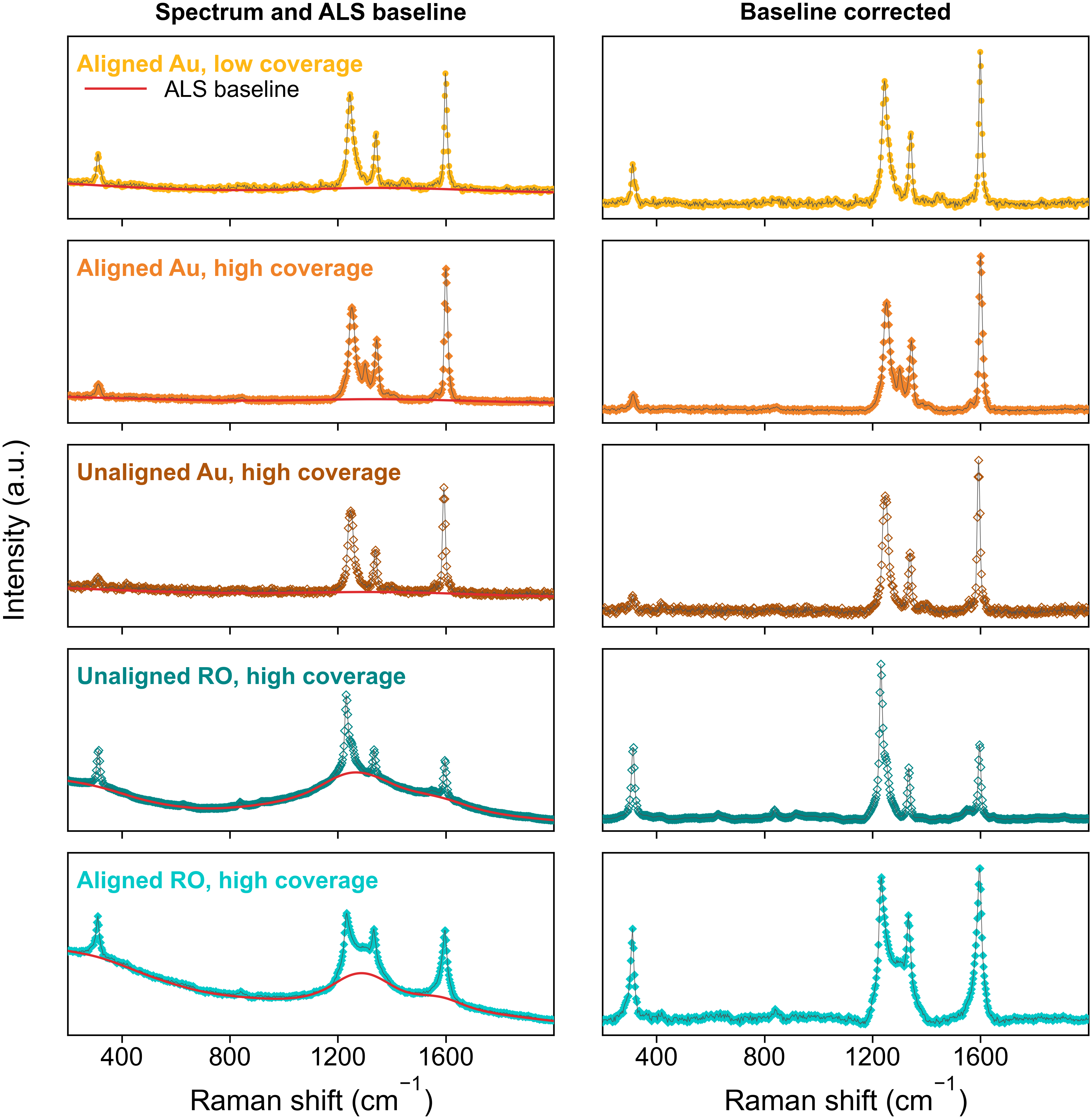}
  \caption{%
  ALS background correction for the five sample configurations. Left column: spike-cleaned spectrum
  together with the smooth ALS baseline (red). Right column: the corresponding
  baseline-corrected spectrum. From top to bottom the rows show \emph{Aligned Au, low-coverage};
  \emph{Aligned Au, high-coverage}; \emph{Unaligned Au, high-coverage};
  \emph{Unaligned RO, high-coverage}; and \emph{Aligned RO, high-coverage}. The correction
  removes the broad background without subtracting
  the narrow Raman bands.}
  \label{fig:ALS}
\end{figure}

\begin{figure}[p]
  \centering
  \includegraphics[width=0.98\textwidth]{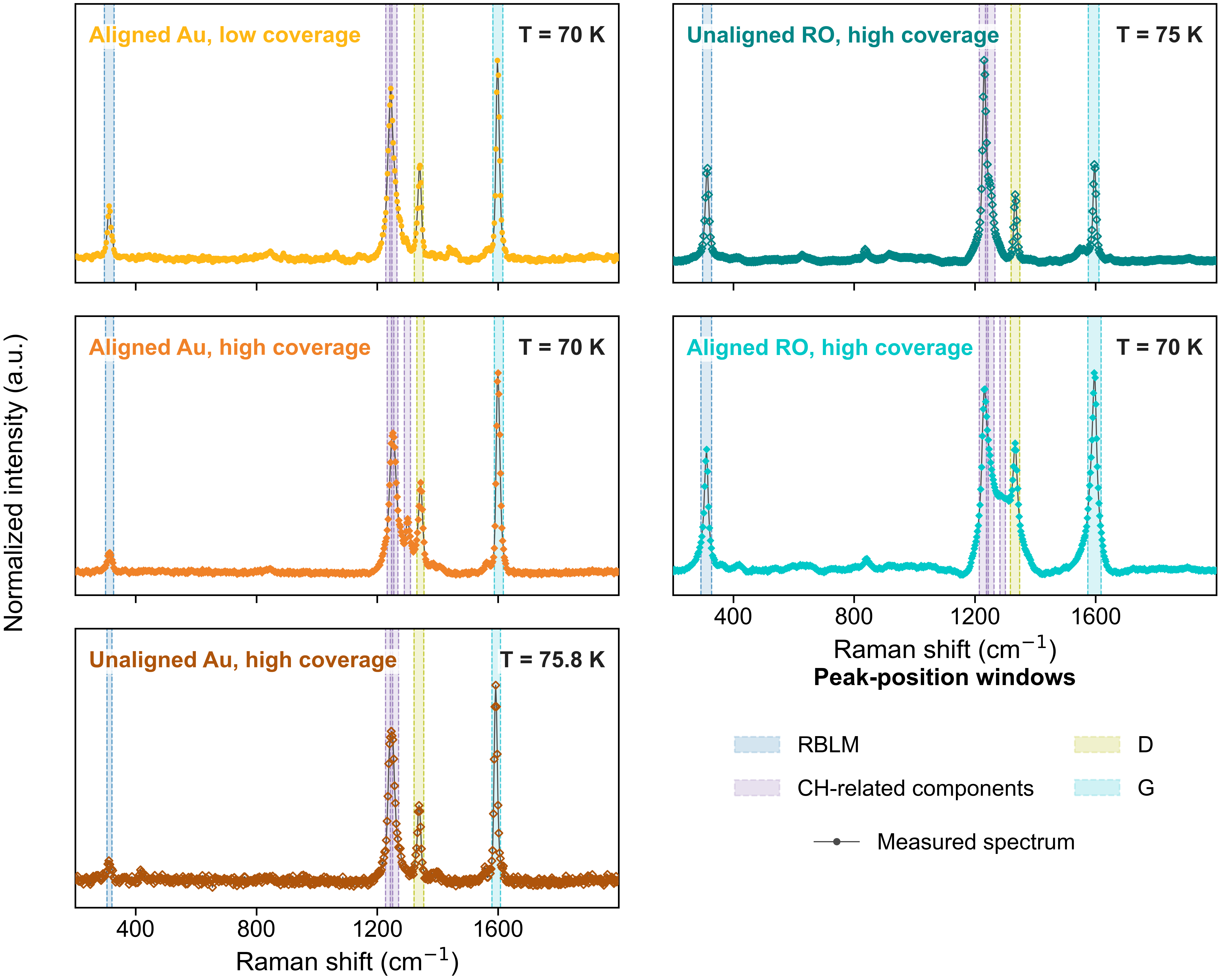}
  \caption{%
  Peak-position windows used for representative fits at $T=70$~K for
  \emph{Aligned Au, low-coverage}; \emph{Aligned Au, high-coverage}; and
  \emph{Aligned RO, high-coverage}; at $T=75$~K for \emph{Unaligned RO, high-coverage};
  and at $T=75.8$~K for \emph{Unaligned Au, high-coverage}. The left column shows the three
  Au-supported configurations, and the right column shows the two Raman-optimised configurations.
  Symbols show baseline-corrected spectra normalised to their respective maxima solely
  for display. Shaded regions and dashed boundaries indicate the final fitted centre-position
  bounds for the RBLM, CH-related, $D$, and $G$ components, where present. These parameter bounds
  are distinct from the broader residual-weighting intervals used in the least-squares objective.}
  \label{fig:fitting_windows}
\end{figure}

Figure~\ref{fig:ALS} shows the procedure for every sample configuration. The slowly varying background
is comparatively weak for the Au-supported samples and is stronger and more curved for the
transferred RO samples. Baseline subtraction produces spectra that fluctuate around zero away
from the Raman bands.

\FloatBarrier

\section{Full-window global Lorentzian fits}

The median sampling intervals on the raw Raman-shift axes are approximately
3.21, 2.68 and 2.56~cm$^{-1}$ in the RBLM, $D$ and $G$ windows, respectively.
Fitting uses these measured coordinates. Peak-position precision is estimated
from the residual-bootstrap procedure below; detector sampling alone does
not set its uncertainty.

Each spike-corrected and baseline-corrected spectrum is fitted over the complete Raman window with
a sum of Lorentzian components,
\begin{equation}
  I_{\mathrm{fit}}(\omega)
  = \sum_{j=1}^{N}
  \frac{H_j}{1+4\left[(\omega-\omega_j)/\Gamma_j\right]^2},
  \label{eq:lorentzian_sum}
\end{equation}
where $H_j$, $\omega_j$, and $\Gamma_j$ are the height, centre frequency, and full width at half
maximum (FWHM) of component $j$. All components within one spectrum are optimised
simultaneously. Each spectrum is fitted independently, without imposing common parameters across
temperatures or sample configurations.

The present study focuses on the RBLM, $D$, and $G$ modes. Additional Lorentzian components were included to describe spectral features observed in the intermediate-frequency region ($1210$--$1355~\mathrm{cm}^{-1}$), including partially overlapping peaks and shoulders. These components represent observed spectral structure and were fitted together with the modes of interest as part of the complete line-shape model. Their detailed assignment, physical origin, and temperature dependence are beyond the scope of the present work and require further investigation, particularly for the aligned Raman-optimised sample, where this region exhibits
a broader and more strongly overlapping response. The number of fitted components was selected according to the spectral features resolved for each sample configuration.

Fits were performed over the full $200$--$2000~\mathrm{cm}^{-1}$ interval using sample- and temperature-dependent bounds on peak position and FWHM. Differential evolution, with a population multiplier of 200, a tolerance of $10^{-7}$, and at most $10^{5}$ generations, provided a global-search solution that was subsequently refined by bounded nonlinear least squares. All measured points within the fitting interval were retained in the objective function. Because this interval contains many more low-intensity points between the Raman bands than points defining the peak profiles, an unweighted sum of squared residuals would be dominated by near-baseline regions and could underrepresent discrepancies in the peak line shapes. To balance these contributions, points within predefined broad windows encompassing
the complete fitted peak profiles were assigned a base objective weight of 6.
Within each window, this weight was further scaled according to the local
normalised intensity by a factor ranging from 1 to 2, giving effective objective
weights from 6 to 12 across the peak profile. Points outside these windows retained
unit weight. The weighting windows were defined before optimisation and only
specified which spectral regions received additional weight. Peak positions and
linewidths remained free parameters within their respective fitting bounds. Regularisation terms suppressed numerically degenerate decompositions in the strongly overlapping CH-related region ($1211$--$1271~\mathrm{cm}^{-1}$), where one component could otherwise become almost entirely buried beneath a neighbouring component. Required-peak visibility penalties were also included. The component parameters
remained free to vary within their prescribed bounds. Reported $R^{2}$ and RMSE values were calculated from the unweighted residuals over the complete fitting interval.

For the full-window fits, parameter uncertainties were estimated from 100
residual-bootstrap replicates. For each replicate, residuals were sampled
with replacement and added to the best-fit spectrum. The synthetic spectrum
was then refitted from parameters perturbed by 3\% relative to the optimum.
The reported bootstrap uncertainty is the sample standard deviation of the
successful refits. Residuals are pooled across the fitted window; final bounds,
weights, preprocessing and line shape are held fixed. Peak-emphasis weights control the fitting objective. The error bars describe
fit precision with these choices held fixed; calibration and systematic
line-shape errors are outside their scope.

For configuration III, 15 of the 16 retained RBLM points were subsequently
refined using a local RBLM fit while the remaining fitted spectral components
were held fixed. The local window, 304.772--320.827~cm$^{-1}$, contains six
measured grid points; the 155~K RBLM point retains the full-window fit. Their
uncertainties were estimated from 100 local residual-bootstrap
refits initiated from the accepted local optimum. The $D$- and $G$-mode
parameters retained the full-window fitting procedure described above.
Separately, configuration-III fits at 125, 145, 185, and 195~K were not
retained after fit review; the same temperature selection was applied to the
RBLM, $D$, and $G$ modes.

\FloatBarrier

\begin{figure}[p]
  \centering
  \includegraphics[width=0.98\textwidth]{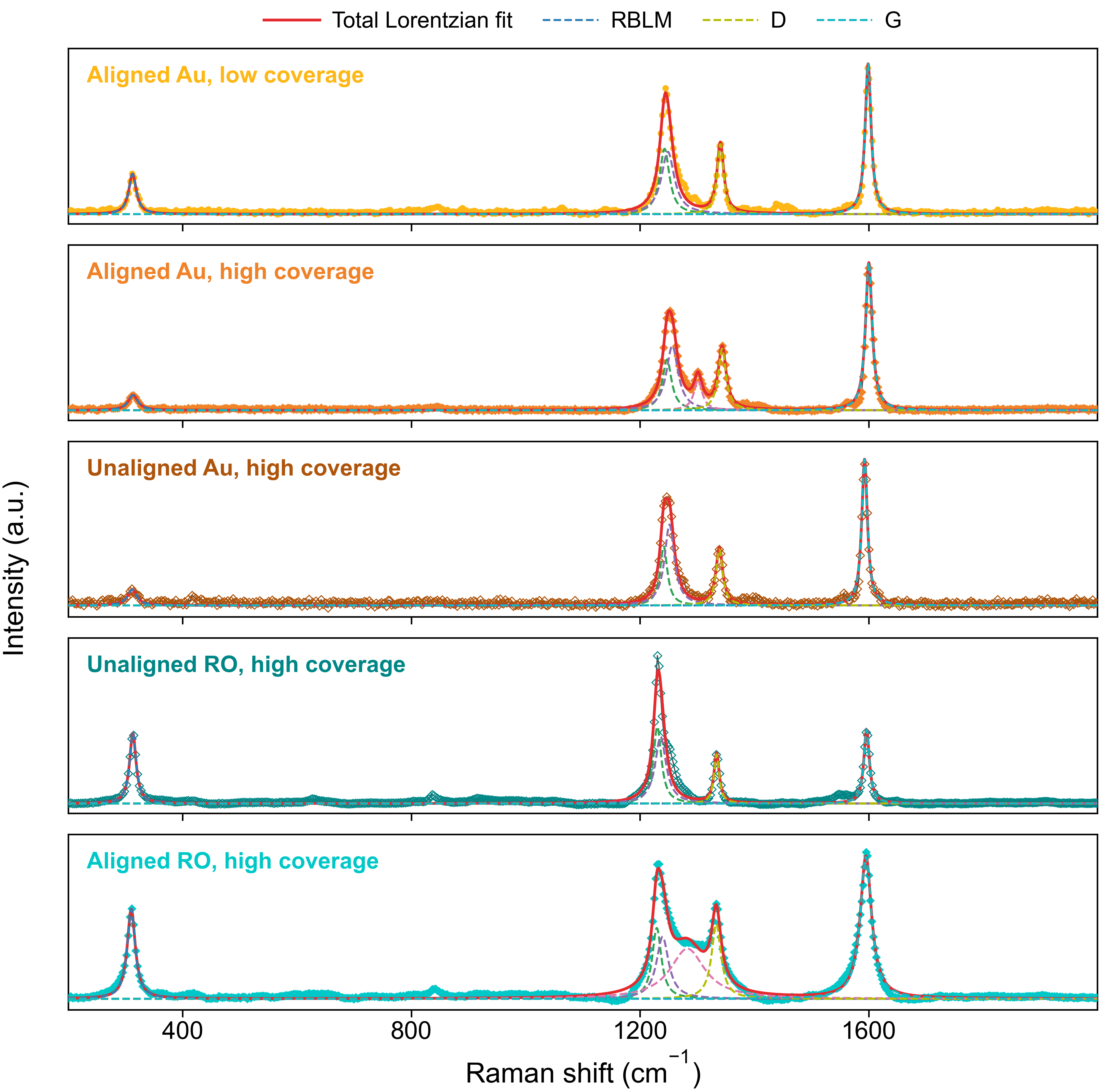}
  \caption{%
  Representative multi-component Lorentzian fits over
  $200$--$2000~\mathrm{cm}^{-1}$ for all five sample configurations. Symbols show the
  baseline-corrected spectra, the red curve shows the total Lorentzian sum, and the dashed curves
  show the individual fitted components. The RBLM, $D$, and $G$ components are identified in the
  legend. Additional intermediate-frequency components between approximately
  $1210$ and $1310~\mathrm{cm}^{-1}$ describe the partially overlapping peaks and shoulders
  observed in this interval. These components are included in the complete line-shape model but
  are not individually assigned in the present study. From top to bottom:
  \emph{Aligned Au, low-coverage}; \emph{Aligned Au, high-coverage};
  \emph{Unaligned Au, high-coverage}; \emph{Unaligned RO, high-coverage}; and
  \emph{Aligned RO, high-coverage}.}
  \label{fig:global_lorentzian}
\end{figure}
\FloatBarrier

\section{Peak position, height, and linewidth from one Lorentzian component}
\label{sec:lorentzian_metrics}

The temperature-dependent data for each Raman mode are taken from the corresponding single Lorentzian component returned by the full-window fit. The fitted centre $\omega_j$, amplitude $H_j$, and linewidth $\Gamma_j$ are used directly as the peak position, peak height, and FWHM, respectively.

\begin{figure}[p]
  \centering
  \includegraphics[width=0.78\textwidth]{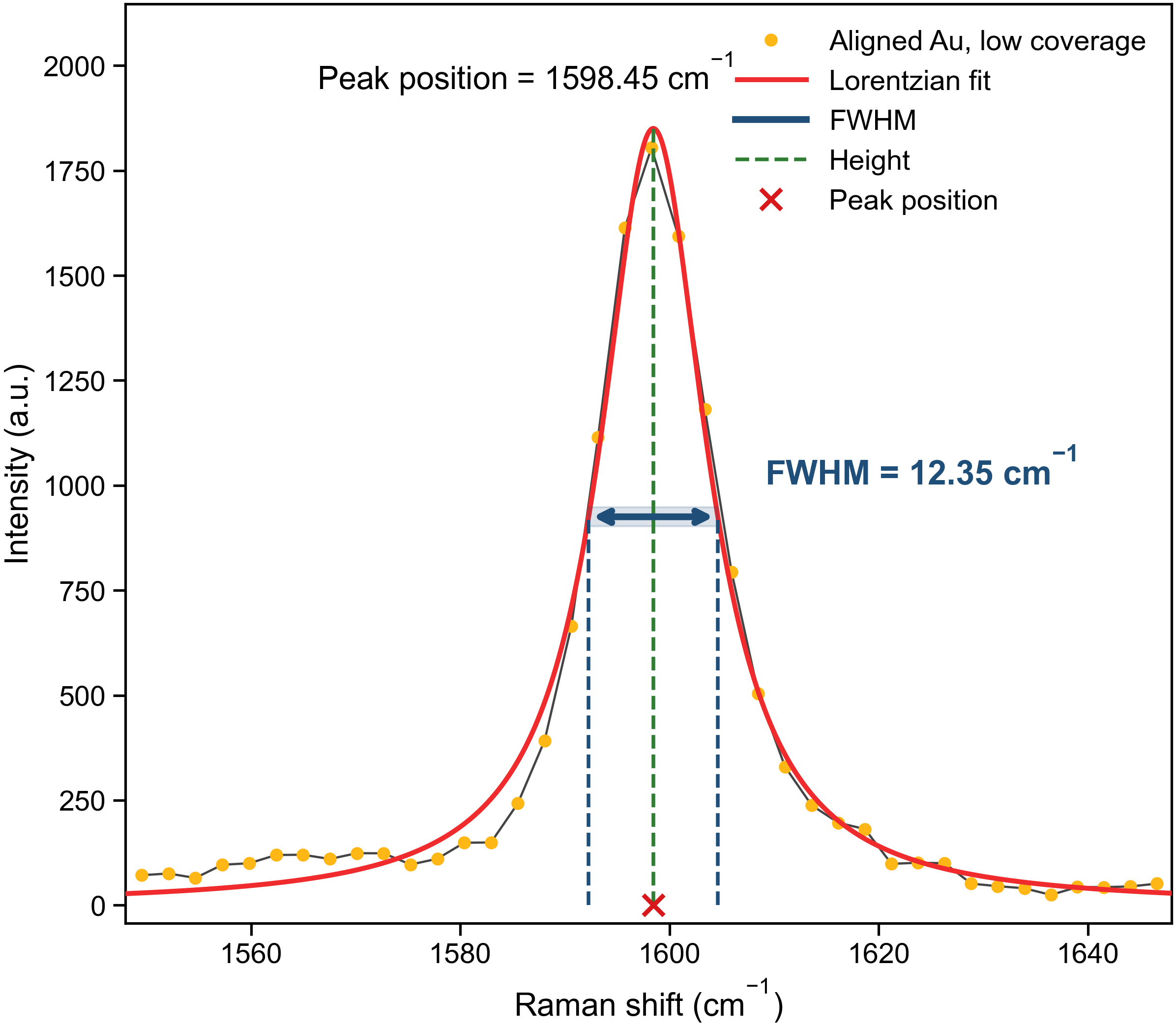}
  \caption{%
  Extraction of the $G$-mode parameters from a single fitted Lorentzian component. Symbols are
  the baseline-corrected data in the $G$-mode window and the red curve is the Lorentzian fit. The
  red cross and green dashed line identify the fitted centre (peak position); the green vertical
  extent represents the peak height; and the blue horizontal arrow, bounded by the two blue
  dashed half-maximum positions, gives the FWHM. For this representative
  \emph{Aligned Au, low-coverage} spectrum, $\omega_G=1598.45$~cm$^{-1}$ and
  $\Gamma_G=12.35$~cm$^{-1}$.}
  \label{fig:lorentzian_metrics}
\end{figure}

\begin{figure}[p]
  \centering
  \includegraphics[width=0.72\textwidth]{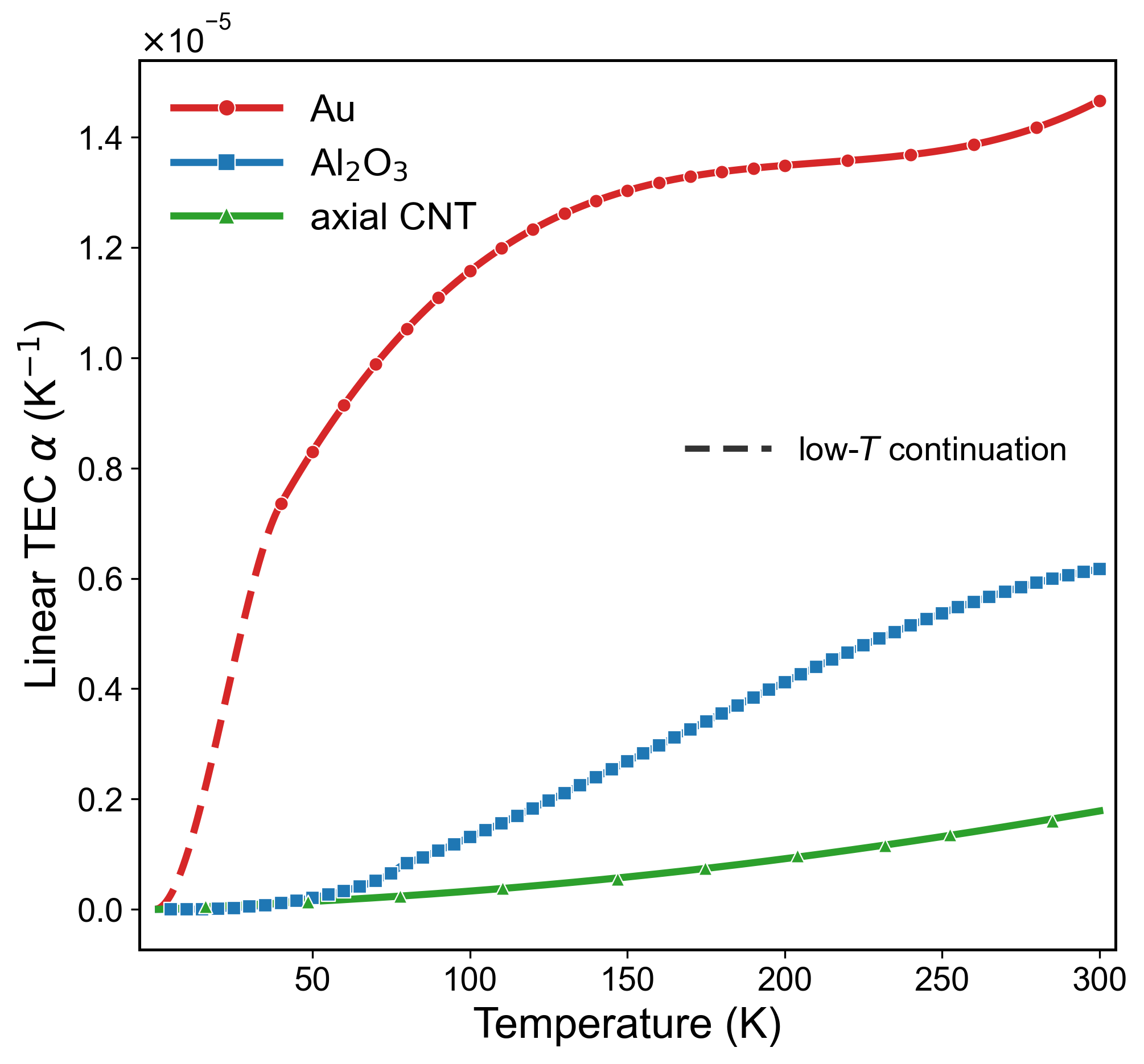}
  \caption{%
  Linear thermal-expansion coefficients $\alpha(T)$ used in the thermomechanical model:
  Au (red), Al$_2$O$_3$ (blue), and the axial single-walled carbon-nanotube coefficient used
  to approximate the longitudinal expansion of 9-AGNRs (green). Au circles show the analytical
  derivative of the Au fit evaluated at the 22 lattice-parameter temperatures; sapphire squares
  are 5-K samples of the continuous NIST-based curve and are not additional measurements; and CNT
  triangles show the digitised source points at $T\leq300$~K. Solid lines show the portions
  derived from the cited data or correlations; dashed lines show only the continuations from
  $\alpha(0)=0$ to the lowest source-supported temperatures (40~K for Au, 5~K for
  Al$_2$O$_3$, and 15.96~K for the axial CNT approximation). The Au and Al$_2$O$_3$
  curves approximate the thermal expansion of the supporting surfaces; no composite
  coefficient was calculated for either complete supporting structure.}
  \label{fig:TEC}
\end{figure}

Figure~\ref{fig:lorentzian_metrics} illustrates these definitions for a representative $G$ mode
of the \emph{Aligned Au, low-coverage} sample. The fitted peak position is 1598.45~cm$^{-1}$ and the
FWHM is 12.35~cm$^{-1}$. The same definitions are applied to every measured temperature and
sample configuration to obtain the peak-position, height, and linewidth series.

\FloatBarrier

\section{Thermal-expansion coefficients used in the thermomechanical model}
\label{sec:thermal_expansion_inputs}

The thermomechanical model requires the linear thermal-expansion coefficient of the supporting
surface, $\alpha_{\mathrm{sub}}(T)$, and that of the nanoribbon along its longitudinal axis,
$\alpha_{\mathrm{GNR}}(T)$. Because a measured longitudinal coefficient for 9-AGNRs is not
available over the experimental temperature range, the calculated axial coefficient of a
single-walled carbon nanotube is used as a surrogate,
$\alpha_{\mathrm{GNR}}(T)\equiv\alpha_{\mathrm{CNT}}(T)$
\cite{Jiang2009ThermalExpansion}.
The implemented mismatch strain is
\begin{equation}
  \varepsilon_{\mathrm{mis}}(T)
  = \int_{0}^{T}
  \left[\alpha_{\mathrm{sub}}(T')-\alpha_{\mathrm{CNT}}(T')\right] \, \mathrm{d}T'.
  \label{eq:mismatch}
\end{equation}
The lower limit is a computational reference fixed at 0~K. Because the Raman measurements begin
near 70~K, the portions below the lowest source-supported temperatures are mathematical
continuations required only to evaluate Eq.~\eqref{eq:mismatch}; they are not additional
measurements.

The same continuously differentiable continuation was used for Au and the axial-CNT surrogate.
For a boundary temperature $T_0$, boundary value $f_0=f(T_0)$, boundary slope
$s_0=f'(T_0)$, and $x=T/T_0$, the continuation is
\begin{equation}
  \mathcal{H}(T;T_0,f_0,s_0)
  = f_0\left(3x^2-2x^3\right)+T_0s_0\left(x^3-x^2\right).
  \label{eq:TEC_Hermite}
\end{equation}
It satisfies $\mathcal{H}(0)=\mathcal{H}'(0)=0$ and matches both $f$ and $f'$ at $T_0$.

For Au, 22 reported lattice parameters between 40 and 300~K were used
\cite{Pamato2018GoldExpansion}.
Linear interpolation gives $a(293~\mathrm{K})=4.0780945$~\AA. The relative expansion
$p_{\mathrm{Au}}(T)=[a(T)-a(293~\mathrm{K})]/a(293~\mathrm{K})$ was fitted by unweighted
least squares to $p_{\mathrm{Au}}(T)=\sum_{k=0}^{4}c_kT^k$, with
\begin{align}
  (c_0,c_1,c_2) &= (-3.358803750394251\times10^{-3},
                    2.394618924216096\times10^{-6},
                    7.495154640183310\times10^{-8}), \nonumber\\
  (c_3,c_4) &= (-2.298896643684569\times10^{-10},
                 2.719719156192072\times10^{-13}).
  \label{eq:Au_TEC_coefficients}
\end{align}
The model uses $\alpha_{\mathrm{Au}}(T)=p'_{\mathrm{Au}}(T)$ from 40 to 300~K and
Eq.~\eqref{eq:TEC_Hermite} below 40~K, with $T_0=40$~K and
$f=p'_{\mathrm{Au}}$. The polynomial is therefore not extrapolated below its fitted range.

For Al$_2$O$_3$, two expressions were taken from the same NIST Sapphire Cryogenic Material
Properties database
\cite{NISTSapphireCryogenic}.
The direct low-temperature coefficient correlation is
\begin{equation}
  \alpha_{\mathrm{L}}(T)
  =10^{-6}\,10^{\sum_{j=0}^{8}a_j[\log_{10}(T)]^j}\ \mathrm{K}^{-1},
  \label{eq:sapphire_low_TEC}
\end{equation}
with
\begin{equation}
  (a_0,\ldots,a_8)=(10.97236,-97.23540,240.2436,-294.9933,195.9244,
  -66.89247,9.19921,0,0).
\end{equation}
The higher-temperature NIST linear-expansion correlation is
$E(T)=10^{-5}\sum_{j=0}^{4}b_jT^j$, with
\begin{equation}
  (b_0,\ldots,b_4)=(-78.850,-2.2346\times10^{-2},1.0185\times10^{-4},
  5.5594\times10^{-6},-8.5422\times10^{-9}),
\end{equation}
and $\alpha_{\mathrm{H}}(T)=E'(T)$. The implemented sapphire coefficient is
\begin{equation}
\alpha_{\mathrm{Al_2O_3}}(T)=
\begin{cases}
  \alpha_{\mathrm{L}}(5)(T/5)^3, & 0\leq T<5~\mathrm{K},\\
  \alpha_{\mathrm{L}}(T), & 5\leq T\leq75~\mathrm{K},\\
  [1-w(u)]\alpha_{\mathrm{L}}(T)+w(u)\alpha_{\mathrm{H}}(T),
    & 75<T<80~\mathrm{K},\\
  \alpha_{\mathrm{H}}(T), & 80\leq T\leq300~\mathrm{K},
\end{cases}
\label{eq:sapphire_piecewise}
\end{equation}
where $u=(T-75~\mathrm{K})/(5~\mathrm{K})$ and $w(u)=u^2(3-2u)$. The smoothstep blend
matches the selected NIST branch and its slope at each end of the 75--80~K interval. The
$T^3$ branch enforces $\alpha(0)=0$ and value continuity at 5~K; unlike
Eq.~\eqref{eq:TEC_Hermite}, it does not impose derivative matching at 5~K.

For the axial-CNT surrogate \cite{Jiang2009ThermalExpansion}, 20 digitised values between 15.96348 and 498.47226~K from a
representative axial-CTE curve in the cited calculation were fitted by unweighted least squares to
$q_{\mathrm{CNT}}(T)=\sum_{k=0}^{4}d_kT^k$, after conversion from
$10^{-6}~\mathrm{K}^{-1}$ to $\mathrm{K}^{-1}$. The coefficients are
\begin{align}
  (d_0,d_1,d_2) &= (1.022158819436424\times10^{-8},
                    2.230194753343012\times10^{-9},
                    6.774688676315011\times10^{-12}), \nonumber\\
  (d_3,d_4) &= (3.441184858543569\times10^{-14},
                -5.362688265083937\times10^{-17}).
  \label{eq:CNT_TEC_coefficients}
\end{align}
The model uses $\alpha_{\mathrm{CNT}}(T)=q_{\mathrm{CNT}}(T)$ above the lowest digitised
temperature and Eq.~\eqref{eq:TEC_Hermite} below it, with $T_0=15.96348$~K and
$f=q_{\mathrm{CNT}}$.

No additive offset or multiplicative scale factor was applied after these fits and correlations
were defined. The three functions were tabulated from 0 to 300~K in 0.1~K increments. During
model fitting they were linearly sampled on an integer-kelvin grid, their difference was
integrated from 0~K by the cumulative trapezoidal rule, and the resulting integral was linearly
interpolated to each measured temperature. For the unaligned sample series containing a 305~K
point, each coefficient was held at its 300~K endpoint between 300 and 305~K. The TEC curves were
held fixed during model fitting; uncertainties associated with the source data,
digitisation, polynomial fits, and effective-support approximation were not
propagated into the fitted-parameter covariance.

Equation~\eqref{eq:mismatch} assumes homogeneous strain transfer between the nanoribbon and its
support. The bulk Au curve represents both the single-crystal Au(788) surface used for the aligned
ribbons and the 200~nm Au film used for the unaligned Au-supported ribbons. The Raman-optimised
substrate consists of a 40~nm ALD-grown Al$_2$O$_3$ layer above an approximately 80~nm metal layer
on SiO$_2$/Si
\cite{Overbeck2019OptimizedSubstrates};
for both Raman-optimised configurations, the bulk sapphire curve is used to
approximate the in-plane thermal expansion of the Al$_2$O$_3$ surface layer,
so $\alpha_{\mathrm{sub}}(T)=\alpha_{\mathrm{Al_2O_3}}(T)$. No thickness-
or elasticity-weighted composite expansion was calculated. The fitted
$\gamma_{\parallel}$ values should therefore be interpreted as effective
coupling parameters within this uniform strain-transfer model, not as unique microscopic
properties of the thin-film or multilayer interfaces.

At 300~K, the adopted coefficients are
$\alpha_{\mathrm{Au}}=1.4668\times10^{-5}~\mathrm{K}^{-1}$,
$\alpha_{\mathrm{Al_2O_3}}=6.1724\times10^{-6}~\mathrm{K}^{-1}$, and
$\alpha_{\mathrm{CNT}}=1.7837\times10^{-6}~\mathrm{K}^{-1}$. Over the experimental interval,
Au has the largest coefficient, Al$_2$O$_3$ is intermediate, and the axial-CNT surrogate is the
smallest.

The instantaneous coefficient difference at 300~K is
$\alpha_{\mathrm{Au}}-\alpha_{\mathrm{CNT}}
=1.2884\times10^{-5}~\mathrm{K}^{-1}$ for Au and
$\alpha_{\mathrm{Al_2O_3}}-\alpha_{\mathrm{CNT}}
=4.3887\times10^{-6}~\mathrm{K}^{-1}$ for Al$_2$O$_3$; their ratio is
2.94. The accumulated mismatch relevant to the common model-comparison
interval is instead
\begin{equation}
  \Delta\varepsilon_{\mathrm{mis}}^{80\rightarrow290}
  =\int_{80~\mathrm{K}}^{290~\mathrm{K}}
  \left[\alpha_{\mathrm{sub}}(T)-
  \alpha_{\mathrm{CNT}}(T)\right]\,\mathrm{d}T .
  \label{eq:mismatch_80_290}
\end{equation}
Evaluation using the adopted thermal-expansion curves gives
$2.5688\times10^{-3}$, or $0.25688$\%, for Au--CNT and
$5.7771\times10^{-4}$, or $0.057771$\%, for Al$_2$O$_3$--CNT. Their ratio
is 4.446, reported as 4.45. This
integrated ratio differs from the instantaneous 300~K ratio because the
coefficient differences vary with temperature. The 4.45 ratio describes
the accumulated strain increment between 80 and 290~K used for comparison
with the Raman peak-position changes over the same interval. Under the
homogeneous, no-slip assumption, these mismatch increments are treated as
axial strain increments transferred to the ribbons; they are not direct
strain measurements.

\FloatBarrier

\section{$G$-region line-shape residuals for the aligned Au samples}

\begin{figure}[p]
  \centering
  \includegraphics[width=0.98\textwidth,height=0.82\textheight,keepaspectratio]{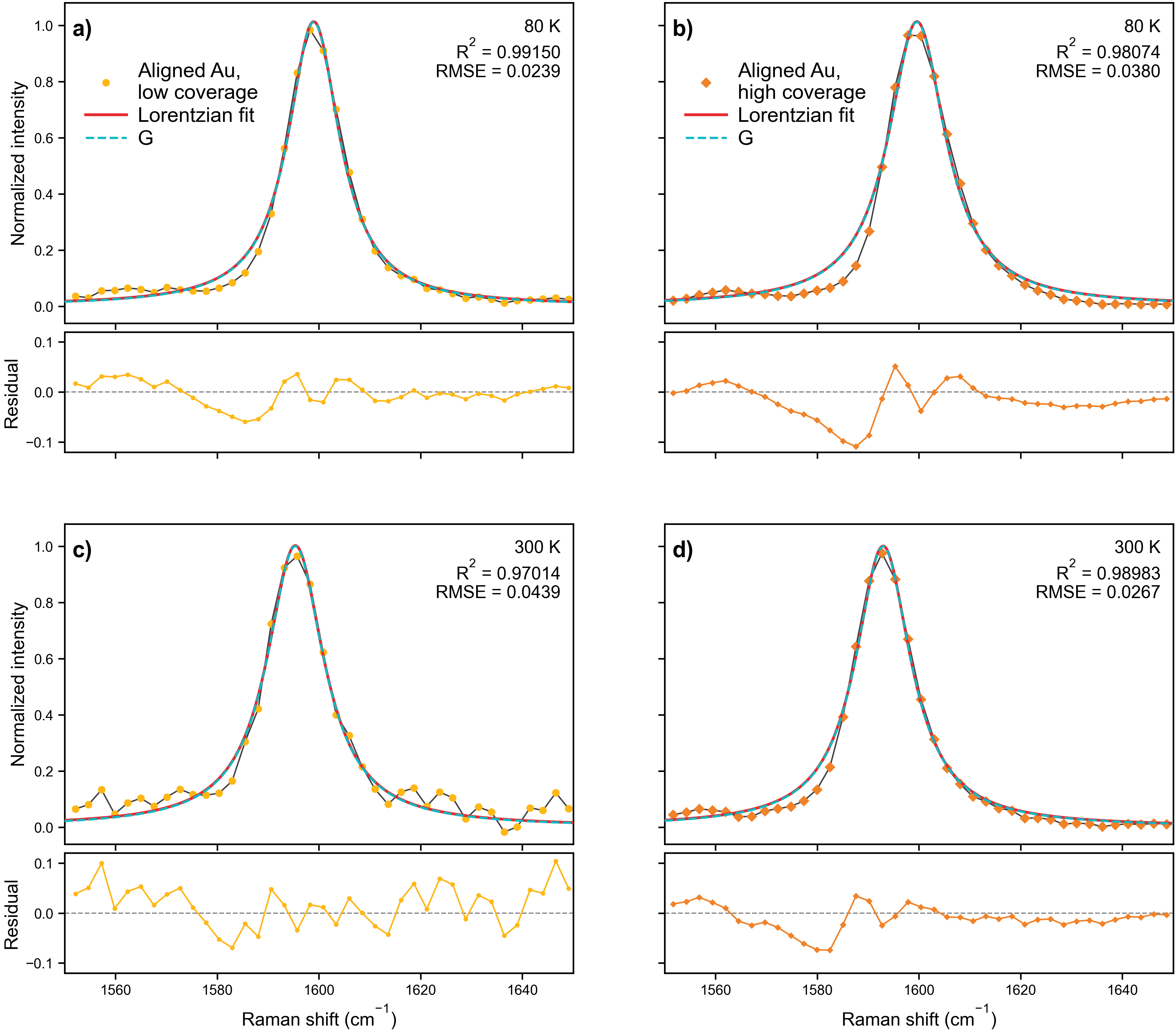}
  \caption{%
  Representative normalised $G$-region spectra, global multi-component Lorentzian fits, fitted
  $G$ components, and residuals for the aligned Au samples. Red curves show the total fitted
  spectra and cyan dashed curves show the fitted $G$ components. Panels (a) and (c) show the
  low-coverage sample at 80 and 300~K,
  respectively; panels (b) and (d) show the high-coverage sample at the same temperatures.
  The fit quality is reported in each panel by $R^2$ and RMSE. At 80~K, the low-coverage
  profile is more nearly symmetric than the high-coverage profile. By 300~K, the high-coverage
  profile is more symmetric and its residual structure is reduced.}
  \label{fig:G_residuals}
\end{figure}

To examine how well the global multi-component Lorentzian model reproduces the $G$-region line
shape for the two aligned Au coverages, Fig.~\ref{fig:G_residuals} compares representative fits at 80 and
300~K. At 80~K, the low-coverage spectrum is more nearly symmetric and gives smaller,
less-structured residuals than the high-coverage spectrum. The latter shows a pronounced
low-frequency-side deviation from the Lorentzian profile. At 300~K, the high-coverage line shape
is more symmetric and its fit residuals are reduced. The low-coverage residuals remain centred
near zero but show greater point-to-point scatter. The temperature evolution is therefore most
apparent for the high-coverage sample; the comparison is not evidence that both coverages follow
the same monotonic change in fit quality.

\FloatBarrier

\section{Thermal cycling and decomposition of the frequency shifts}

\begin{figure}[p]
  \centering
  \includegraphics[width=0.98\textwidth,height=0.76\textheight,keepaspectratio]{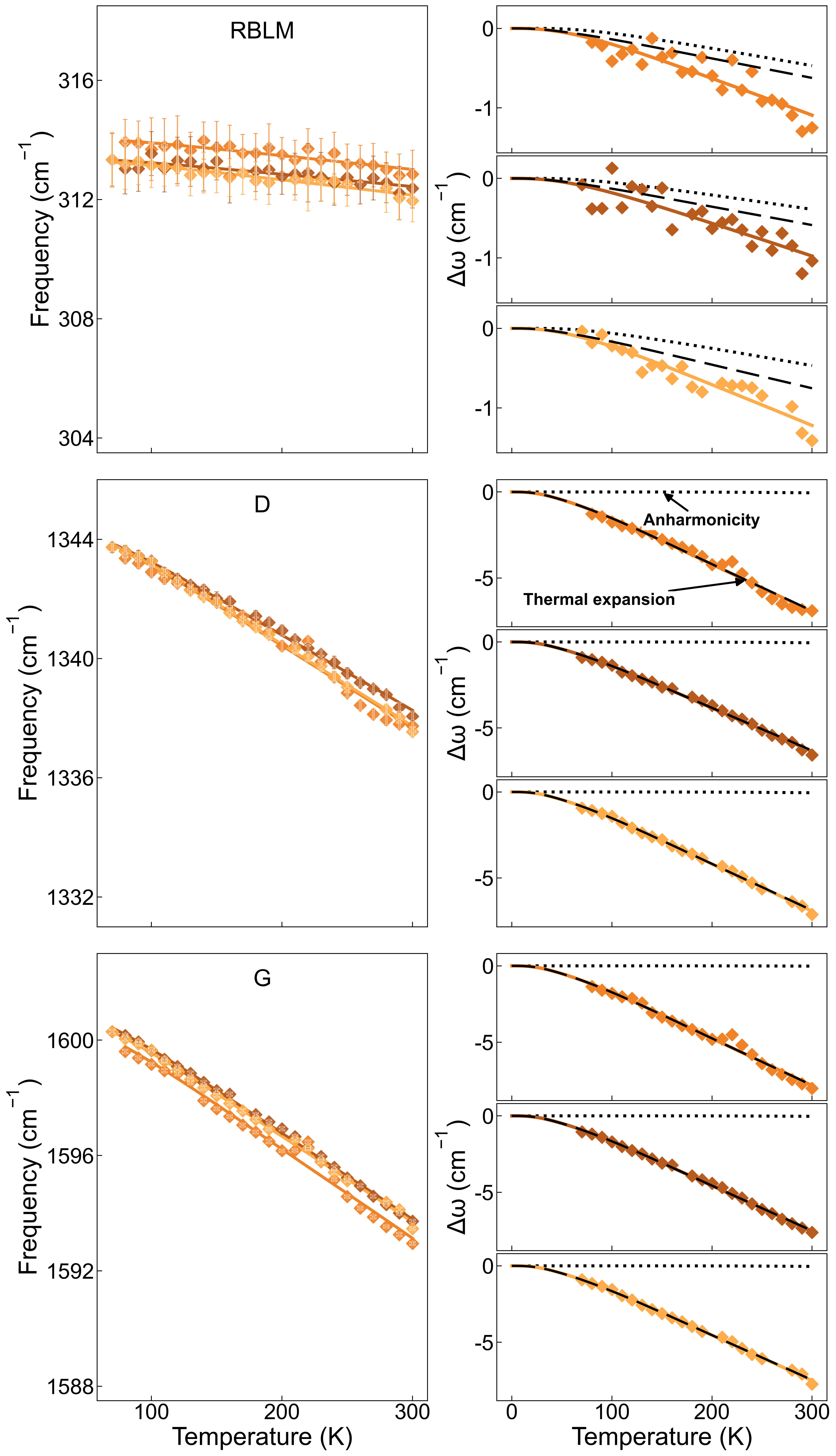}
  \caption{%
  Temperature-dependent Raman peak positions and model decomposition for the
  \emph{Aligned Au, high-coverage}
  sample during successive heating, cooling, and reheating. Left column: measured
  RBLM, $D$-mode, and $G$-mode peak positions for the three thermal paths,
  with vertical bars showing their one-standard-deviation residual-bootstrap
  uncertainties and coloured curves showing the corresponding total models.
  Right column: the Raman peak-position shifts for heating, cooling, and
  reheating shown separately. Coloured solid curves give the total modelled shift, black dashed
  curves give the thermoelastic contribution, and black dotted curves give the zero-K-referenced
  effective Klemens-type contribution of Eq.~\eqref{eq:frequency_anharmonic_referenced}; arrows in the first
  $D$-mode panel identify the two model components. The three paths reproduce
  the same overall $D$- and $G$-mode redshift, although resolved path-dependent
  offsets remain; one measured cycle does not establish a reproducible
  thermodynamic hysteresis loop. The fixed-reference decomposition assigns most $D/G$ shift to the
  thermoelastic term under the stated reference and TEC assumptions. Right-hand
  ordinates are absolute shifts in cm$^{-1}$.}
  \label{fig:thermal_cycles}
\end{figure}

The high-coverage aligned Au sample was measured through successive heating,
cooling, and reheating paths. Figure~\ref{fig:thermal_cycles} compares the
RBLM, $D$, and $G$ Raman peak positions from all three paths. The three paths
retain the same overall $D$- and $G$-mode redshift with increasing
temperature, and their independently fitted slopes are compared in
Table~\ref{tab:thermal_path_comparison}. They do not, however, retrace
exactly at every shared temperature. At 280~K, the Heating~2 $G$-mode peak
position is $0.85~\mathrm{cm}^{-1}$ higher than during Heating~1, compared
with a combined residual-bootstrap uncertainty of
$0.06~\mathrm{cm}^{-1}$. At 270~K, the Cooling~1 $D$-mode peak position is
$0.85~\mathrm{cm}^{-1}$ higher than during Heating~1, compared with
$0.23~\mathrm{cm}^{-1}$. These offsets exceed the peak-fitting
uncertainties, but because only one complete cycle was measured, the data do
not establish a reproducible thermodynamic hysteresis loop.

For each thermal path, the unreferenced model frequency is
\begin{equation}
  \omega(T)=\omega_0
  +\Delta\omega_{\mathrm{TE}}(T)
  +\Delta\omega_{\mathrm{anh}}(T),
  \label{eq:frequency_model}
\end{equation}
where $\Delta\omega_{\mathrm{TE}}$ is the thermoelastic contribution
generated by the mismatch strain in Eq.~\eqref{eq:mismatch}, and
$\Delta\omega_{\mathrm{anh}}$ is an effective Klemens-type term that
represents the temperature dependence expected for a lowest-order symmetric
three-phonon process,
\begin{equation}
  \begin{aligned}
  \Delta\omega_{\mathrm{anh}}(T)
    &= A_3\left[1+2n_{\mathrm{B}}\!\left(\frac{\omega_0}{2},T\right)\right],\\
  n_{\mathrm{B}}(\widetilde{\nu},T)
    &= \left[\exp\!\left(\frac{hc\widetilde{\nu}}{k_{\mathrm{B}}T}\right)-1\right]^{-1}.
  \end{aligned}
  \label{eq:frequency_anharmonic}
\end{equation}
Here $\omega_0$ is the extrapolated reference Raman wavenumber of the parent
RBLM, $D$, or $G$ mode, whereas $\widetilde{\nu}$ is the daughter-phonon
wavenumber entering the Bose--Einstein occupation. In the symmetric
three-phonon Klemens process, each daughter phonon has half the parent-mode
energy, so $\widetilde{\nu}=\omega_0/2$. Both wavenumbers are expressed in
cm$^{-1}$, with $c$ expressed in cm~s$^{-1}$. Because $\omega_0$ is an
empirical extrapolated intercept rather than an independently calculated
harmonic frequency, $A_3$ is an effective signed coefficient in cm$^{-1}$
and is not interpreted as a uniquely isolated cubic self-energy. A negative
$A_3$ produces a Klemens-type softening as temperature increases, whereas
a positive $A_3$ produces hardening. With the limiting value
$n_{\mathrm{B}}(\omega_0/2,0)=0$, the unreferenced expression contains the
zero-point term $A_3$.
The component displayed as a temperature-induced shift is referenced to the full model at 0~K
and is therefore
\begin{equation}
  \Delta\omega_{\mathrm{anh}}^{(0)}(T)
  \equiv \Delta\omega_{\mathrm{anh}}(T)-\Delta\omega_{\mathrm{anh}}(0)
  =2A_3n_{\mathrm{B}}\!\left(\frac{\omega_0}{2},T\right).
  \label{eq:frequency_anharmonic_referenced}
\end{equation}
Accordingly, in the right-hand panels of Fig.~\ref{fig:thermal_cycles}, the coloured solid curve
is the zero-K-referenced total shift, the black dashed curve is the thermoelastic contribution,
and the black dotted curve is $\Delta\omega_{\mathrm{anh}}^{(0)}$. The fixed-reference model assigns most $D$- and $G$-mode shifts
to the thermoelastic term for this sample; this decomposition is reference-model
dependent. For the much smaller RBLM shift, the two
contributions are more comparable and the relative interpretation is correspondingly more
sensitive to scatter.

\FloatBarrier

\section{Linewidth trends and the three-phonon model}

\begin{figure}[p]
  \centering
  \includegraphics[width=0.98\textwidth,height=0.82\textheight,keepaspectratio]{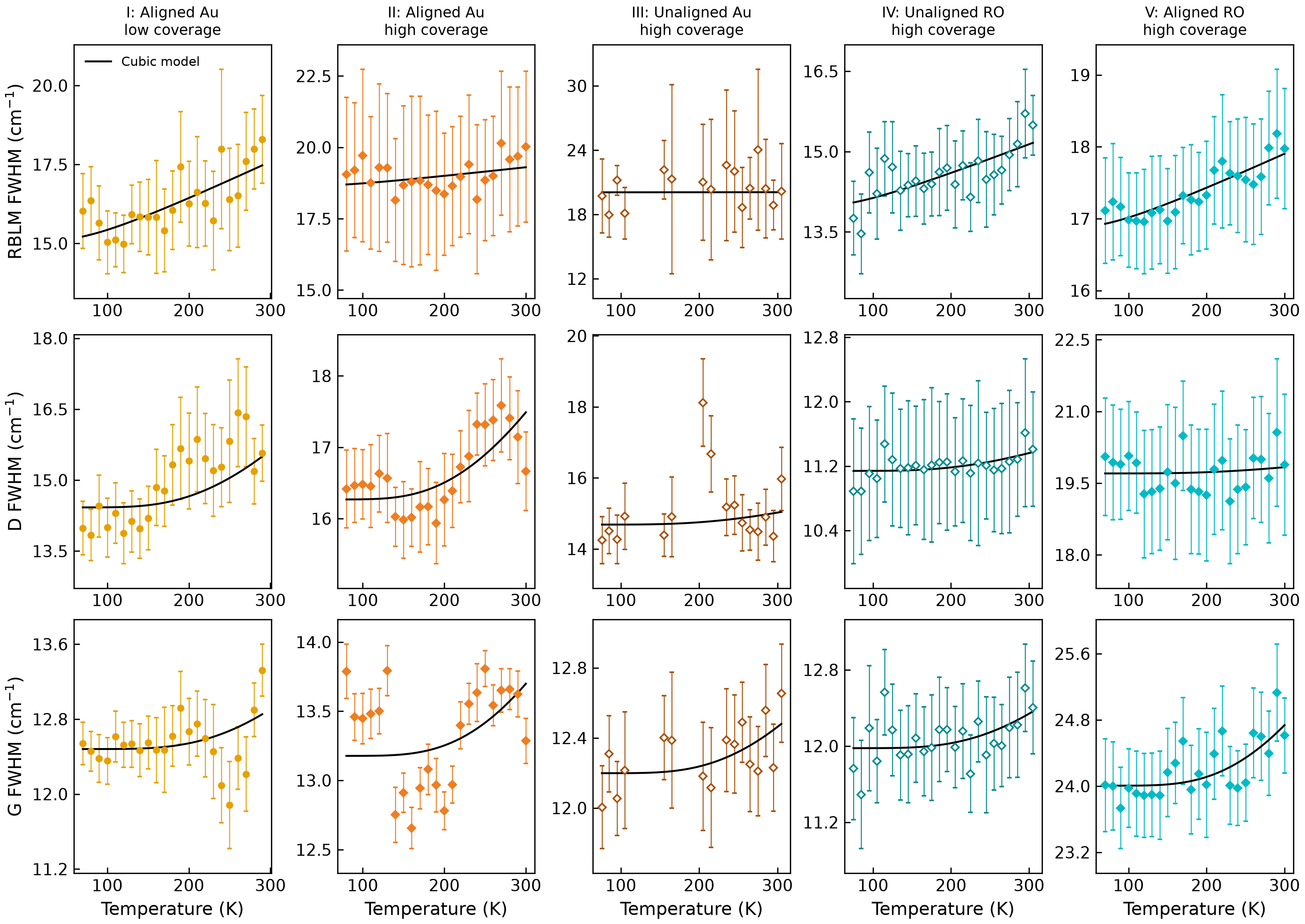}
  \caption{%
  Temperature dependence of the Lorentzian FWHM for the RBLM (top row), $D$ mode (middle row),
  and $G$ mode (bottom row). Columns, from left to right:
  \emph{Aligned Au, low-coverage}; \emph{Aligned Au, high-coverage};
  \emph{Unaligned Au, high-coverage}; \emph{Unaligned RO, high-coverage}; and
  \emph{Aligned RO, high-coverage}. The displayed paths are Heating for I,
  III, and IV and Heating~1 for II and V. Symbols show the fitted linewidths
at each temperature, vertical bars show their
  one-standard-deviation residual-bootstrap uncertainties, and black curves
  are the three-phonon fits of Eq.~\eqref{eq:Gamma_anh}. Each axis has its own numerical scale. The
  non-monotonic structure, most clearly resolved in the $G$ linewidth of the
  \emph{Aligned Au, high-coverage} sample, is not fully
  described by the single-channel model.}
  \label{fig:linewidths}
\end{figure}

\begin{figure}[p]
  \centering
  \includegraphics[width=0.98\textwidth,height=0.82\textheight,keepaspectratio]{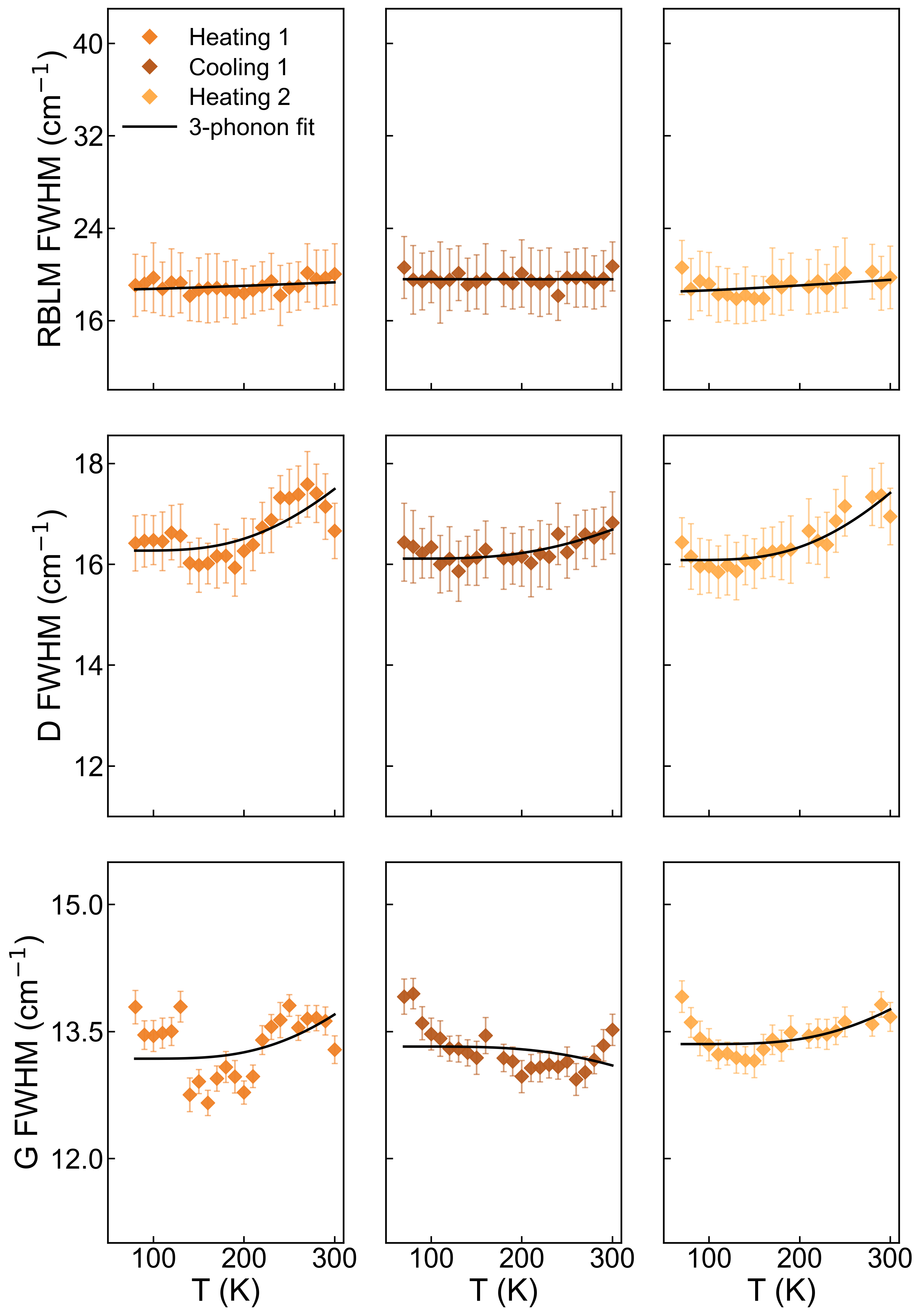}
  \caption{%
  Temperature dependence of the Lorentzian FWHM during thermal cycling of the
  \emph{Aligned Au, high-coverage} sample. Rows show the RBLM (top), $D$ mode (middle), and $G$ mode (bottom);
  columns, from left to right, show Heating~1, Cooling~1, and Heating~2. Colours identify the
  thermal paths as indicated by the in-panel legend. Symbols show the fitted linewidths at each temperature, vertical bars show their
  one-standard-deviation residual-bootstrap uncertainties, and black curves are the three-phonon fits of
  Eq.~\eqref{eq:Gamma_anh}. The $G$ peak narrows and then broadens during Heating~1. Cooling and
  reheating show related non-monotonic trends. The $D$ changes are smaller
  relative to their fitting uncertainties. The curves show the monotonic
  three-phonon baseline.}
  \label{fig:linewidth_cycles}
\end{figure}

The Lorentzian FWHM values extracted as described in
Sec.~\ref{sec:lorentzian_metrics} were analysed as a function of
temperature for the RBLM, $D$, and $G$ modes. The black curves in
Figs.~\ref{fig:linewidths} and~\ref{fig:linewidth_cycles} are fits to the
Klemens-type three-phonon linewidth model
\begin{equation}
  \Gamma_j(T)=\Gamma_{0,j}+A_{\Gamma,j}
  \left[1+2n_{\mathrm{B}}\!\left(
  \frac{\widetilde{\nu}_{j,\mathrm{LT}}}{2},T\right)\right],
  \label{eq:Gamma_anh}
\end{equation}
where $n_{\mathrm{B}}$ is defined in Eq.~\eqref{eq:frequency_anharmonic},
and $\widetilde{\nu}_{j,\mathrm{LT}}$ is the low-temperature reference
wavenumber of the parent Raman mode $j$. For each configuration and thermal
path, $\widetilde{\nu}_{j,\mathrm{LT}}$ was fixed to the median fitted peak
centre at the lowest measured temperature available for that series. In the
symmetric three-phonon Klemens channel, each daughter phonon is assigned half
the parent-mode energy, so the Bose--Einstein occupation is evaluated at
$\widetilde{\nu}_{j,\mathrm{LT}}/2$. The reference wavenumber was fixed
because its measured temperature variation is small relative to its
absolute value and the linewidth data do not independently constrain it.
Thus, only $\Gamma_{0,j}$ and $A_{\Gamma,j}$ were fitted. Here
$\Gamma_{0,j}$ represents the temperature-independent residual contribution,
including static and instrumental broadening. Because the three-phonon term
includes its zero-point contribution,
$\Gamma_j(0)=\Gamma_{0,j}+A_{\Gamma,j}$. Only this lowest-order
three-phonon channel was fitted over each measured temperature series
(70--305~K overall). A conventional positive four-phonon term cannot supply initial narrowing.
The cubic fit provides a monotonic anharmonic baseline. Other scattering
processes may also contribute to the measured width.

For each fitted spectrum, the linewidth uncertainty is the sample
standard deviation of the linewidth values obtained from the 100 successful
residual-bootstrap refits described in Section~S3. These one-standard-deviation
uncertainties are plotted as the vertical bars and were used as relative
weights in the nonlinear least-squares linewidth fits.

Figure~\ref{fig:linewidths} shows mode rows and configuration columns,
with a numerical scale on each axis. In II Heating~1, the $G$ peak narrows
towards intermediate temperatures and broadens again at higher temperatures.
Cooling and reheating show related non-monotonic trends with different depths
and positions (Fig.~\ref{fig:linewidth_cycles}). The corresponding $D$ changes
are smaller relative to their fitting uncertainties. The measured widths can
include spatial variation and unresolved spectral structure as well as
homogeneous phonon broadening.

\FloatBarrier

\FloatBarrier

\section{Inter-ribbon spacing and illustrative edge-contact geometry}

The backbone-to-backbone pitch $p$ is the transverse distance between ribbon
centres. For backbone width $w$, the carbon-edge gap is
$d_{\mathrm{edge-edge}}=p-w$. Taking $w\approx0.95$--$1.0$~nm, a collinear
illustration using two $1.09$~\AA{} C--H bonds and a $2.2$--$2.4$~\AA{}
H-centre contact separation gives a carbon-edge gap of $0.44$--$0.46$~nm
and a pitch of about $1.4$--$1.5$~nm. This construction assumes transverse,
collinear bonds and opposing hydrogen centres; bond projection, longitudinal
registry, relative ribbon height and the stepped surface can change it.
It is not a universal hard-wall lower bound.

Three rows per $3.8$--$3.9$~nm terrace instead give an average row period
of $1.27$--$1.30$~nm. Occupancy and terrace width do not determine each
nearest-neighbour gap. These two illustrative ranges do not overlap, and no
measured uncertainty is available to claim numerical agreement. The structural
evidence establishes denser packing at high coverage
\cite{Darawish2025PrecursorCoverage}, while the local separations and forces
are not measured here. The calculations in Ref.~\cite{Chang2014EdgeDecorated}
concern interacting edges during ribbon curvature and provide only a qualitative
analogy for repulsion, not a quantitative model of adjacent flat ribbons on Au.

\clearpage
\section{Parameters of the thermomechanical model}

The frequency model in Eq.~\eqref{eq:frequency_model} combines the effective
Klemens-type contribution defined in Eq.~\eqref{eq:frequency_anharmonic}
with the shift generated by
substrate--ribbon thermal-expansion mismatch. The thermoelastic contribution is
\begin{align}
  \Delta\omega_{\mathrm{TE}}(T)
  &= \omega_0\left\{\exp\!\left[-\gamma_{\parallel}
     \mathcal{I}(T)\right]-1\right\}, \\
  \mathcal{I}(T)
  &= \int_0^T\left[\alpha_{\mathrm{sub}}(T')-
     \alpha_{\mathrm{GNR}}(T')\right]\,\mathrm{d}T'.
  \label{eq:thermomechanical_terms}
\end{align}
The fixed reference $\omega_0$ is obtained before fitting: by an unweighted linear extrapolation
of the points at $T\leq110$~K for the
$D$ and $G$ modes, and by an all-temperature linear extrapolation for the RBLM. The free model
parameters are $A_3$ and the effective coupling parameter
$\gamma_{\parallel}$. As shown in Eq.~\eqref{eq:frequency_anharmonic_referenced}, the plotted
frequency changes subtract the full model at 0~K, so no temperature-independent constant is
counted as a temperature-induced shift.

Table~\ref{tab:model_params} summarises the fixed $\omega_0$ values and the fitted parameters.
Two contributions to the
uncertainty of $\omega_0$ are evaluated: propagation of the measured peak-position uncertainties
through the extrapolation and the uncertainty associated with the residual scatter of that
extrapolation. The larger estimate is reported. The uncertainties of $A_3$ and $\gamma_{\parallel}$ are one-standard-error
estimates from the covariance matrix of the unweighted nonlinear fit, with
$\omega_0$ and the TEC curves held fixed. The reference-frequency uncertainty
is reported separately and is not propagated into these parameter errors. Uncertainty
in the TEC curves and in the bulk-sapphire and axial-CNT approximations is not propagated. For several RBLM and RO series, the total frequency shift is comparable to the
point-to-point scatter. The corresponding lower $R^2$ values and larger parameter uncertainties
indicate weaker parameter identifiability.

\begin{landscape}
\begin{table}[p]
  \centering
  \footnotesize
  \setlength{\tabcolsep}{4.5pt}
  \caption{%
  Thermomechanical quantities used in and obtained from the temperature-dependent peak positions.
  Frequencies, $A_3$, and RMSE are in cm$^{-1}$. Values following $\pm$ are one-standard-error
  estimates, with the distinct definitions given in the text. In the adopted sign convention,
  $A_3<0$ denotes temperature-induced softening and $A_3>0$ denotes hardening. The thermal path identifies whether the temperature series was measured during
  heating or cooling. For the cyclic measurement, Heating~1 denotes the initial heating path.}
  \label{tab:model_params}
  \begin{tabular}{@{}llc r@{\,$\pm$\,}l r@{\,$\pm$\,}l r@{\,$\pm$\,}l rr@{}}
    \toprule
    Mode & Sample configuration & Thermal path & \multicolumn{2}{c}{$\omega_0$}
      & \multicolumn{2}{c}{$A_3$} & \multicolumn{2}{c}{$\gamma_{\parallel}$}
      & RMSE & $R^2$ \\
    \midrule
    RBLM & \emph{Aligned Au, low-coverage}    & Heating   & 312.76 & 0.21 & -0.206 & 0.081 & 0.578 & 0.264 & 0.129 & 0.793 \\
    RBLM & \emph{Aligned Au, high-coverage}   & Heating 1 & 314.37 & 0.51 & -0.264 & 0.099 & 0.623 & 0.312 & 0.146 & 0.808 \\
    RBLM & \emph{Unaligned Au, high-coverage} & Heating   & 311.70 & 0.49 & -0.073 & 0.137 & 0.205 & 0.421 & 0.176 & 0.261 \\
    RBLM & \emph{Unaligned RO, high-coverage} & Heating   & 312.95 & 0.15 & -0.106 & 0.029 & 0.632 & 0.573 & 0.110 & 0.419 \\
    RBLM & \emph{Aligned RO, high-coverage}   & Heating 1 & 310.45 & 0.14 & -0.241 & 0.024 & 0.972 & 0.502 & 0.093 & 0.813 \\
    \addlinespace
    $D$  & \emph{Aligned Au, low-coverage}    & Heating   & 1341.53 & 0.52 & -0.539 & 0.054 & 0.820 & 0.022 & 0.105 & 0.987 \\
    $D$  & \emph{Aligned Au, high-coverage}   & Heating 1 & 1345.27 & 0.76 & -0.628 & 0.134 & 1.619 & 0.051 & 0.245 & 0.982 \\
    $D$  & \emph{Unaligned Au, high-coverage} & Heating   & 1339.12 & 1.00 & -0.269 & 0.155 & 0.382 & 0.054 & 0.243 & 0.806 \\
    $D$  & \emph{Unaligned RO, high-coverage} & Heating   & 1334.01 & 1.08 &  0.194 & 0.054 & 0.415 & 0.130 & 0.160 & 0.308 \\
    $D$  & \emph{Aligned RO, high-coverage}   & Heating 1 & 1333.59 & 0.80 & -0.205 & 0.029 & 1.031 & 0.073 & 0.088 & 0.908 \\
    \addlinespace
    $G$  & \emph{Aligned Au, low-coverage}    & Heating   & 1600.25 & 0.42 & -0.770 & 0.039 & 0.767 & 0.013 & 0.077 & 0.994 \\
    $G$  & \emph{Aligned Au, high-coverage}   & Heating 1 & 1601.39 & 0.25 & -0.418 & 0.136 & 1.537 & 0.044 & 0.254 & 0.984 \\
    $G$  & \emph{Unaligned Au, high-coverage} & Heating   & 1594.52 & 1.02 & -1.271 & 0.123 & 0.468 & 0.035 & 0.195 & 0.933 \\
    $G$  & \emph{Unaligned RO, high-coverage} & Heating   & 1596.49 & 1.11 &  0.050 & 0.064 & 0.416 & 0.127 & 0.190 & 0.333 \\
    $G$  & \emph{Aligned RO, high-coverage}   & Heating 1 & 1595.21 & 0.50 &  0.030 & 0.024 & 0.832 & 0.050 & 0.072 & 0.929 \\
    \bottomrule
  \end{tabular}
\end{table}
\end{landscape}
\FloatBarrier

\section{Linear temperature slopes}

Table~\ref{tab:slope_comparison} compares two linear slopes over each measured temperature
series. The data slope, $b_{\mathrm{data}}$, is obtained by fitting a straight line directly
through the measured Raman peak positions. To obtain the fitted-curve slope,
$b_{\mathrm{model}}$, the fitted thermomechanical curve is evaluated at the same experimental
temperatures and a straight line is fitted through those values. Their
relative difference is
\begin{equation}
  \mathrm{Rel.\ diff.}
  =100\,\frac{b_{\mathrm{data}}-b_{\mathrm{model}}}{b_{\mathrm{model}}}.
  \label{eq:slope_difference}
\end{equation}
The two central estimates agree within 5.1\% for 13 of the 15 series;
this comparison is an internal consistency check on the same data. The approximately 11\%
differences for the $D$ and $G$ modes of \emph{Unaligned RO, high-coverage} occur because their total
shifts are very small relative to the point-to-point scatter.

\begin{landscape}
\begin{table}[p]
  \centering
  \small
  \setlength{\tabcolsep}{7pt}
  \caption{%
  Linear slopes from measured peak positions and fitted thermomechanical curves
  (cm$^{-1}$K$^{-1}$) are displayed as rounded central estimates describing each measured
  temperature range and thermal path; relative differences use the unrounded slopes.
  Uncertainties of the spectral peak positions and thermomechanical fit parameters are
  reported separately, with no additional uncertainty propagation implied here.}
  \label{tab:slope_comparison}
  \begin{tabular}{@{}llcrrr@{}}
    \toprule
    Mode & Sample configuration & Thermal path & $b_{\mathrm{model}}$ & $b_{\mathrm{data}}$ & Rel. diff. (\%) \\
    \midrule
    RBLM & \emph{Aligned Au, low-coverage}    & Heating   & -0.0038 & -0.0038 &  0.4 \\
    RBLM & \emph{Aligned Au, high-coverage}   & Heating 1 & -0.0044 & -0.0045 &  1.2 \\
    RBLM & \emph{Unaligned Au, high-coverage} & Heating   & -0.0013 & -0.0014 &  0.4 \\
    RBLM & \emph{Unaligned RO, high-coverage} & Heating   & -0.0014 & -0.0013 & -3.7 \\
    RBLM & \emph{Aligned RO, high-coverage}   & Heating 1 & -0.0027 & -0.0028 &  5.1 \\
    \addlinespace
    $D$  & \emph{Aligned Au, low-coverage}    & Heating   & -0.0137 & -0.0137 & 0.0 \\
    $D$  & \emph{Aligned Au, high-coverage}   & Heating 1 & -0.0271 & -0.0271 &  0.1 \\
    $D$  & \emph{Unaligned Au, high-coverage} & Heating   & -0.0064 & -0.0064 &  0.3 \\
    $D$  & \emph{Unaligned RO, high-coverage} & Heating   & -0.0015 & -0.0014 & -11.4 \\
    $D$  & \emph{Aligned RO, high-coverage}   & Heating 1 & -0.0039 & -0.0040 &  3.0 \\
    \addlinespace
    $G$  & \emph{Aligned Au, low-coverage}    & Heating   & -0.0152 & -0.0152 &  0.0 \\
    $G$  & \emph{Aligned Au, high-coverage}   & Heating 1 & -0.0305 & -0.0305 &  0.0 \\
    $G$  & \emph{Unaligned Au, high-coverage} & Heating   & -0.0094 & -0.0094 & 0.0 \\
    $G$  & \emph{Unaligned RO, high-coverage} & Heating   & -0.0019 & -0.0017 & -10.8 \\
    $G$  & \emph{Aligned RO, high-coverage}   & Heating 1 & -0.0037 & -0.0038 &  2.0 \\
    \bottomrule
  \end{tabular}
\end{table}
\end{landscape}
\FloatBarrier

\section{Additive decomposition of the modelled frequency shift}

To compare all sample configurations over one interval without mixing reference temperatures, the decomposition
in Table~\ref{tab:shift_decomposition} is evaluated from 80 to 290~K, which lies within the common
measurement window. For each channel $x\in\{\mathrm{TE},\mathrm{anh}\}$,
\begin{equation}
  \delta\omega_x = \Delta\omega_x(290~\mathrm{K})
  -\Delta\omega_x(80~\mathrm{K}),
  \qquad
  \delta\omega_{\mathrm{tot}}
  =\delta\omega_{\mathrm{TE}}+\delta\omega_{\mathrm{anh}}.
  \label{eq:additive_decomposition}
\end{equation}
The signed shares are $100\,\delta\omega_x/\delta\omega_{\mathrm{tot}}$ and therefore sum
exactly to 100\%. The IV $D$ row provides a concrete example:
$-0.320+0.015=-0.305~\mathrm{cm}^{-1}$. Here the positive fitted Klemens-type change is a small
hardening contribution that partially offsets the larger negative \emph{thermoelastic}
softening. Division by the smaller net redshift consequently gives a thermoelastic share above
100\% and a negative effective Klemens-type share. These are signed component ratios, not absolute
fractions or a percentage error. The positive $G$-mode Klemens-type changes in IV and V have the
same interpretation.

The values in Table~\ref{tab:shift_decomposition} are central model estimates evaluated
from the fitted parameters. Uncertainties in the fixed reference frequencies,
thermal-expansion inputs and mechanical approximations are not propagated into this
decomposition; the displayed digits therefore describe numerical model output and should
not be interpreted as the independently resolved experimental precision of the corresponding
Raman-frequency changes.

\begin{landscape}
\begin{table}[p]
  \centering
  \small
  \setlength{\tabcolsep}{6pt}
  \caption{%
  Additive decomposition of the modelled 80--290~K frequency shift. Frequency changes are in
  cm$^{-1}$; negative changes denote softening and positive changes denote hardening. Shares are
  signed percentages of the total modelled shift, so a small hardening term that opposes a larger
  softening term appears as a negative share while the softening share can exceed 100\%. The decomposition uses
  the parameters reported in Table~\ref{tab:model_params}. Component and total changes were
  calculated at full precision and rounded independently for display; minor last-digit
  discrepancies in displayed sums therefore reflect rounding. The column ``Anh.'' denotes
  the zero-K-referenced effective Klemens-type contribution.}
  \label{tab:shift_decomposition}
  \begin{tabular}{@{}llcrrrrr@{}}
    \toprule
    Mode & Sample configuration & Thermal path & $\delta\omega_{\mathrm{TE}}$
      & $\delta\omega_{\mathrm{anh}}$ & $\delta\omega_{\mathrm{tot}}$
      & TE (\%) & Anh. (\%) \\
    \midrule
    RBLM & \emph{Aligned Au, low-coverage}    & Heating   & -0.464 & -0.325 & -0.789 &  58.8 &  41.2 \\
    RBLM & \emph{Aligned Au, high-coverage}   & Heating 1 & -0.503 & -0.413 & -0.916 &  54.9 &  45.1 \\
    RBLM & \emph{Unaligned Au, high-coverage} & Heating   & -0.164 & -0.116 & -0.279 &  58.6 &  41.4 \\
    RBLM & \emph{Unaligned RO, high-coverage} & Heating   & -0.114 & -0.167 & -0.281 &  40.6 &  59.4 \\
    RBLM & \emph{Aligned RO, high-coverage}   & Heating 1 & -0.174 & -0.384 & -0.558 &  31.2 &  68.8 \\
    \addlinespace
    $D$  & \emph{Aligned Au, low-coverage}    & Heating   & -2.822 & -0.040 & -2.862 &  98.6 &   1.4 \\
    $D$  & \emph{Aligned Au, high-coverage}   & Heating 1 & -5.580 & -0.046 & -5.627 &  99.2 &   0.8 \\
    $D$  & \emph{Unaligned Au, high-coverage} & Heating   & -1.314 & -0.020 & -1.334 &  98.5 &   1.5 \\
    $D$  & \emph{Unaligned RO, high-coverage} & Heating   & -0.320 &  0.015 & -0.305 & 104.8 &  -4.8 \\
    $D$  & \emph{Aligned RO, high-coverage}   & Heating 1 & -0.794 & -0.016 & -0.810 &  98.1 &   1.9 \\
    \addlinespace
    $G$  & \emph{Aligned Au, low-coverage}    & Heating   & -3.149 & -0.030 & -3.178 &  99.1 &   0.9 \\
    $G$  & \emph{Aligned Au, high-coverage}   & Heating 1 & -6.306 & -0.016 & -6.322 &  99.7 &   0.3 \\
    $G$  & \emph{Unaligned Au, high-coverage} & Heating   & -1.915 & -0.050 & -1.965 &  97.5 &   2.5 \\
    $G$  & \emph{Unaligned RO, high-coverage} & Heating   & -0.384 &  0.002 & -0.382 & 100.5 &  -0.5 \\
    $G$  & \emph{Aligned RO, high-coverage}   & Heating 1 & -0.766 &  0.001 & -0.765 & 100.2 &  -0.2 \\
    \bottomrule
  \end{tabular}
\end{table}
\end{landscape}
\FloatBarrier

\section{Thermal-path comparison for configuration II}

Heating~1, Cooling~1, and Heating~2 of configuration II
(\emph{Aligned Au, high-coverage}) were analysed independently using the
same thermomechanical model. Table~\ref{tab:thermal_path_comparison}
compares the reference and fitted quantities obtained for the three paths.

All six $D$- and $G$-mode data slopes are negative. Relative to Heating~1,
the magnitude of the $D$-mode slope differs by 9.2\% for Cooling~1 and
1.2\% for Heating~2. The corresponding differences for the $G$ mode are
4.1\% and 4.8\%. For every $D$ and $G$ path, the slope of the fitted curve
agrees with the slope obtained directly from the measured peak positions
within 0.12\%, and the fits give $R^2=0.982$--$0.998$. The three paths
therefore retain the same overall $D$- and $G$-mode redshift with increasing
temperature, although individual fitted parameters and peak positions are
not identical between paths.

The RBLM comparison is less precise. Its data-slope magnitude ranges from
$0.00395$ to $0.00489~\mathrm{cm}^{-1}\mathrm{K}^{-1}$ and its
$R^2$ values range from 0.758 to 0.900. This greater relative variation is
consistent with the smaller RBLM temperature shift and its weaker parameter
identifiability.

\begin{landscape}
\begin{table}[p]
  \centering
  \footnotesize
  \setlength{\tabcolsep}{4pt}
  \caption{%
  Independent path-by-path analysis of configuration II
  (\emph{Aligned Au, high-coverage}). Heating~1 is the initial heating
  path, Cooling~1 is the subsequent cooling path, and Heating~2 is the
  reheating path. Each path was analysed independently using the same
  thermal-expansion functions and fitting procedure. The uncertainty in
  $\omega_0$ is the larger of the propagated peak-position uncertainty and
  the uncertainty associated with the scatter of the extrapolation.
  Uncertainties in $A_3$ and $\gamma_{\parallel}$ are one-standard-error
  estimates from the fit covariance matrix, conditional on the fixed
  $\omega_0$ and thermal-expansion functions. Frequencies, $A_3$, and RMSE
  are in cm$^{-1}$; $b_{\mathrm{data}}$ and $b_{\mathrm{model}}$ are in
  cm$^{-1}$K$^{-1}$; $\gamma_{\parallel}$ and $R^2$ are dimensionless.
  The two slopes are defined as in Table~\ref{tab:slope_comparison}.}
  \label{tab:thermal_path_comparison}
  \begin{tabular}{@{}ll
    r@{\,$\pm$\,}l
    r@{\,$\pm$\,}l
    r@{\,$\pm$\,}l
    rrrr@{}}
    \toprule
    Mode & Thermal path
      & \multicolumn{2}{c}{$\omega_0$}
      & \multicolumn{2}{c}{$A_3$}
      & \multicolumn{2}{c}{$\gamma_{\parallel}$}
      & RMSE & $R^2$ & $b_{\mathrm{data}}$ & $b_{\mathrm{model}}$ \\
    \midrule
    RBLM & Heating~1
      & 314.37 & 0.51 & -0.264 & 0.099 & 0.623 & 0.312
      & 0.146 & 0.808 & -0.00449 & -0.00444 \\
    RBLM & Cooling~1
      & 313.63 & 0.46 & -0.218 & 0.098 & 0.588 & 0.312
      & 0.159 & 0.758 & -0.00395 & -0.00392 \\
    RBLM & Heating~2
      & 313.63 & 0.46 & -0.262 & 0.071 & 0.755 & 0.230
      & 0.113 & 0.900 & -0.00489 & -0.00489 \\
    \addlinespace
    $D$ & Heating~1
      & 1345.27 & 0.76 & -0.628 & 0.134 & 1.619 & 0.051
      & 0.245 & 0.982 & -0.02714 & -0.02711 \\
    $D$ & Cooling~1
      & 1345.22 & 0.48 & -0.594 & 0.052 & 1.478 & 0.020
      & 0.103 & 0.996 & -0.02463 & -0.02462 \\
    $D$ & Heating~2
      & 1345.23 & 0.45 & -0.561 & 0.041 & 1.613 & 0.017
      & 0.080 & 0.998 & -0.02682 & -0.02681 \\
    \addlinespace
    $G$ & Heating~1
      & 1601.39 & 0.25 & -0.418 & 0.136 & 1.537 & 0.044
      & 0.254 & 0.984 & -0.03047 & -0.03045 \\
    $G$ & Cooling~1
      & 1602.07 & 0.19 & -0.722 & 0.043 & 1.479 & 0.014
      & 0.087 & 0.998 & -0.02921 & -0.02921 \\
    $G$ & Heating~2
      & 1602.04 & 0.19 & -0.835 & 0.048 & 1.468 & 0.016
      & 0.094 & 0.998 & -0.02902 & -0.02901 \\
    \bottomrule
  \end{tabular}
\end{table}
\end{landscape}
\FloatBarrier

\section{Post-transfer spectroscopic assessment of configuration V}
\label{sec:configuration_V_transfer_assessment}

Configuration V (\emph{Aligned RO, high-coverage}) was produced by
transferring an aligned high-coverage array from Au(788) to the RO
substrate. Configuration II (\emph{Aligned Au, high-coverage}) provides the
corresponding pre-transfer reference. The characteristic RBLM, $D$, and $G$
bands remain present after transfer, as shown at 100~K in Fig.~1(c) of the
main text, throughout the measured temperature range in Fig.~2 of the main
text, and in the representative global fits of
Fig.~\ref{fig:global_lorentzian}. The persistence of these bands shows that
the principal Raman fingerprint of the 9-AGNR array is retained. The
post-transfer line shapes are not, however, spectroscopically equivalent to
those measured in II. Because Au and the RO substrate have different
optical responses, this assessment is based on spectral resolution and
fitted linewidths rather than absolute Raman intensity.

The difference is clearest in the $D$- and $G$-mode linewidths. At 100~K,
comparison of Cooling~1 for II with Heating~1 for V gives
\[
  \Gamma_D^{\mathrm{II}}=16.34\pm0.61~\mathrm{cm}^{-1},
  \qquad
  \Gamma_D^{\mathrm{V}}=20.07\pm1.14~\mathrm{cm}^{-1},
\]
and
\[
  \Gamma_G^{\mathrm{II}}=13.47\pm0.19~\mathrm{cm}^{-1},
  \qquad
  \Gamma_G^{\mathrm{V}}=23.97\pm0.47~\mathrm{cm}^{-1}.
\]
The $D$ and $G$ linewidths therefore increase by
$3.73\pm1.29~\mathrm{cm}^{-1}$ and
 $10.50\pm0.51~\mathrm{cm}^{-1}$, respectively, where the uncertainty of
each difference is the quadrature sum of the two independent
residual-bootstrap standard deviations. Relative to II, these changes
correspond to increases of 22.8\% for $D$ and 78.0\% for $G$.

At the same
temperature, the three thermal paths of II give only
$15.96$--$16.48~\mathrm{cm}^{-1}$ for $D$ and
$13.34$--$13.47~\mathrm{cm}^{-1}$ for $G$. The II$\rightarrow$V changes
are therefore substantially larger than the path-to-path variation measured
before transfer.

The broadening persists across the complete shared temperature range. Across
the 23 temperatures shared by II Cooling~1 and V Heating~1 between 70 and
300~K, the mean $D$-mode
FWHM increases from $16.28$ to $19.69~\mathrm{cm}^{-1}$, an increase of
20.9\%, while the mean $G$-mode FWHM increases from $13.29$ to
$24.19~\mathrm{cm}^{-1}$, an increase of 82.1\%. Both modes are broader
in V at every shared temperature. The V series is shown in
Fig.~\ref{fig:linewidths}, whereas Cooling~1 of II is shown in
Fig.~\ref{fig:linewidth_cycles}. The
larger uncertainties obtained for the broader V profiles indicate reduced
precision of the fitted linewidths under residual resampling; they do not
by themselves indicate failed optimisation or the appearance of a separate
vibrational mode.

The intermediate CH-related region provides independent evidence of reduced
spectral resolution. In II at 100~K on Cooling~1, the additional structure
near $1300~\mathrm{cm}^{-1}$ is localised by the global fit at
$1300.97\pm0.42~\mathrm{cm}^{-1}$ with a FWHM of
$17.38\pm1.47~\mathrm{cm}^{-1}$. None of the 67 fitted spectra from the
three thermal paths of II places this component at either its centre-position
or linewidth boundary. In V during Heating~1, the corresponding structure forms part of a
much broader and more strongly overlapping CH-related/$D$ envelope. Its
fitted centre reaches the lower permitted bound in all 24 spectra, while
its linewidth reaches the upper bound in 6 of the 24 spectra. At 100~K,
the fit returns the lower centre bound of $1283~\mathrm{cm}^{-1}$ and the
upper linewidth bound of $73.81~\mathrm{cm}^{-1}$. These boundary-limited
values are not interpreted as the physical position and linewidth of a
resolved Raman band. Instead, the broad auxiliary Lorentzian visible in the
bottom row of Fig.~\ref{fig:global_lorentzian} represents an unresolved
intermediate-frequency envelope. Likewise, the weak low-wavenumber
shoulder or asymmetry adjacent to the $G$ band in II is not separately
resolved from the broadened $G$ response in V. Because this shoulder was
not fitted as an independent component, it is treated only as a qualitative
line-shape observation.

The RBLM does not exhibit the same broadening. At 100~K, comparison of
Cooling~1 for II with Heating~1 for V gives an FWHM change from
$19.79\pm2.22~\mathrm{cm}^{-1}$ in II to
$16.98\pm0.66~\mathrm{cm}^{-1}$ in V. The difference,
$-2.81\pm2.32~\mathrm{cm}^{-1}$, is not resolved within the combined
uncertainty and is therefore not used to assess the transfer outcome. The
preserved RBLM, $D$, and $G$ bands, together with the absence of uniform
broadening across all three modes, argue against wholesale loss of the
9-AGNR Raman fingerprint.

The observed changes establish systematic $D$- and $G$-mode broadening and
loss of independently resolved structure in the intermediate-frequency
region after transfer. They do not determine the microscopic origin.
Possible contributions include inhomogeneous broadening from distributions
of local strain, ribbon--substrate contact, or adsorption registry,
together with unresolved overlapping Raman components or changes in
homogeneous phonon dephasing. Broadening alone does not establish the
formation of atomic defects, and no defect assignment is made here.

Accordingly, V is described as a \emph{spectroscopically non-equivalent
post-transfer sample}, denoting its changed Raman line shape. It is not a quantitative measurement of transfer yield or
efficiency. Because V is not spectroscopically equivalent to its
pre-transfer reference II, the II$\rightarrow$V comparison cannot isolate
the effect of replacing Au with the RO substrate. Configuration V is
therefore retained in the figures and parameter tables as a diagnostic
configuration but is not used in the principal mechanistic comparisons.

\clearpage

\bibliographystyle{elsarticle-harv}
\bibliography{references}